\documentclass[prl,reprint,aps,superscriptaddress,longbibliography]{revtex4-2}
\usepackage{graphicx}% Include figure files
\usepackage{dcolumn}% Align table columns on decimal point
\usepackage{bm}% bold math
\usepackage{braket}
\usepackage{amsmath}
\usepackage{amssymb}
\usepackage{feynmp}
\usepackage{graphics}
\usepackage{enumerate}
\usepackage[dvipsnames]{xcolor}

\usepackage{comment}
\usepackage[normalem]{ulem} % sout 
\usepackage[hidelinks]{hyperref}% add hypertext capabilities

\everymath{\displaystyle}

\newcommand{\JS}[1]{{\color{violet}{#1}}}

\newcommand{\bmA}{{\bm A}}
\newcommand{\bmr}{{\bm r}}
\newcommand{\bmq}{{\bm q}}
\newcommand{\bmk}{{\bm k}}
\newcommand{\bmu}{{\bm u}}
\newcommand{\bmR}{{\bm R}}
\newcommand{\bmG}{{\bm G}}

\begin{document}

\title{
Effects of Berry curvature on ideal fractional Chern insulator many-body gaps}
%Berry curvature effects on the stability of fractional Chern insulators in ideal bands

\author{Jingtian Shi}
\affiliation{Materials Science Division, Argonne National Laboratory, Lemont, Illinois 60439, USA}

\author{Jennifer Cano}
\affiliation{Center for Computational Quantum Physics, Flatiron Institute, New York, New York 10010, USA}
\affiliation{Department of Physics and Astronomy, Stony Brook University, Stony Brook, New York 11794, USA}

\author{Nicol\'as Morales-Dur\'an}
\email{nmoralesduran@flatironinstitute.org}
\affiliation{Center for Computational Quantum Physics, Flatiron Institute, New York, New York 10010, USA}

\date{\today}

\begin{abstract}
We investigate the many-body ground states in a family of fractionally-filled bands where the Berry curvature fluctuations can be tuned while maintaining ideal quantum geometry. We numerically find that the neutral gap of the fractional Chern insulator (FCI) ground state decreases as the Berry curvature becomes less homogeneous, ultimately driving an instability to a charge density wave. We further extend our analysis to bands perturbed away from the ideal limit and give examples where a less ideal band geometry results in a more stable FCI phase. To explain our findings, we apply the single mode approximation to the ground state wave functions of the ideal band, from which we obtain analytic expressions for the magnetoroton minimum. Finally, we make a connection between our results and experimentally relevant systems where FCIs have been observed.
\end{abstract}

\maketitle

\emph{Introduction ---}
Fractional Chern insulators (FCIs) are the analogs of fractional quantum Hall (FQH) states in the absence of  magnetic field. They were proposed over a decade ago \cite{Wen_FCI,DasSarma_Sun_FCI,Neupert_FCI,Sheng_FCI,Bernevig_Regnault_FCI,Bernevig_Regnault_FCI2}, but their realization remained elusive until recently, when observations in twisted transition metal dichalcogenide (TMD) homobilayers \cite{cai2023signatures, zeng2023thermodynamic, park2023observation, xu2023observation} and rhombohedral graphene multilayers \cite{lu2024fractional} opened a new frontier in the field of topological order \cite{FQAH_Review1,FQAH_Review2}. Follow-up experimental studies have observed rich new phenomenology in both the TMD \cite{Cornell_FQSH1,Cornell_FQSH2,Washington_Higherband,Columbia_Higherbands} and graphene \cite{LongJu_ExtandedHall,choi2024electricfieldcontrolsuperconductivity} platforms, giving rise to several puzzles regarding the nature of moiré FCIs. In this context, it is crucial to obtain a theoretical understanding of how electronic correlations in two-dimensional materials give rise to physics in the absence of a magnetic field that resembles -- but at the same time reveals striking differences from -- the FQH paradigm in Landau levels.

The observed FCIs are often studied theoretically by means of continuum models subject to periodic spatial modulations from the moir\'e superlattice \cite{wu2019topological}. The framework of quantum geometry \cite{roy2014band,jackson2015geometric,wang2021chiral,wang2021exact,wang2023origin} links those models to FQH physics through the notion of {\it ideal bands}, which are exactly flat bands whose wave functions can be mapped to those of the lowest Landau level (LLL) precisely via a spatial modulation. Hamiltonians with ideal bands are appealing because, for a particular form of the electron-electron interactions, analytic many-body ground state wave functions can be obtained by generalizing the Laughlin construction \cite{laughlin1983anomalous,Kivelson_Trugman}. Examples of ideal band models include the chiral model of twisted bilayer graphene \cite{tarnopolsky2019origin} and its multilayer generalizations \cite{wang2022hierarchy,ledwith2022family}, Dirac and spin-$1/2$ Schr\"odinger particles in a periodic magnetic field \cite{aharonov1979ground,dubrovin1980ground,Dong_PeriodicDirac}, the Kapit-Mueller model \cite{Kapit_Mueller} and models with higher band touchings \cite{Sun_QuadraticTouching,Sun_Chern2}. Although real materials are not expected to be described by fine-tuned ideal band models, the experimental discovery of FQH physics over wide ranges of twist angle and displacement field suggests that some moiré FCIs may be approximately described by perturbing away from an ideal band limit. Thus, it is important to understand how deviations from the limits of ideal geometry and constant Berry curvature impact the stability of putative FCI ground states. However, in generic models, finite band width, violations of ideal band geometry, and variations in Berry curvature often appear simultaneously, making it difficult to disentangle their effects on the ground state.
In this work, we gain insight by isolating these effects.

We first investigate the many-body physics of a family of ideal band models, that we refer to as Aharonov-Casher (AC) Hamiltonians. These models were introduced previously to approximate the continuum model for homobilayer TMDs \cite{morales-duran2024magic, AharonovCasher_TMD}, but their many-body physics still remains to be fully explored \cite{Fengcheng_variational}. The AC models have the advantage that the Berry curvature distribution of the topmost band can be tuned while maintaining the ideal quantum geometry and vanishing band width.
This isolates the effect of non-homogeneous Berry curvature on
the FCI that results when the top band is partially filled.
Our exact diagonalization (ED) calculations show that for a realistic Coulomb interaction, the neutral gap of the FCI ground state decreases when Berry curvature fluctuations are increased, despite the band remaining ideal. We also observe that the magnitude of the magnetoroton gap depends on the location of the Berry curvature extrema in the original band. We explain these observations by exploiting the exact solvability of the AC Hamiltonian. Specifically, we apply the single mode approximation (SMA) in the limit of small Berry curvature inhomogeneity to obtain an analytic expression for the magnetoroton minimum, which agrees qualitatively with our numerics. 

Next, we study the many-body physics of band models with small violations from the ideal condition. Surprisingly, we find instances where sacrificing ideal geometry in favor of a more homogeneous Berry curvature increases the stability of the topologically ordered phase. We explain this effect using our SMA framework and relate it to realistic scenarios where FCIs have been experimentally observed. Our results show that the adiabatic approximation \cite{morales-duran2024magic,AharonovCasher_TMD} provides a good description of the continuum model for twisted TMD homobilayers at both the single-particle and many-body levels, and justify the application of numerical and analytical many-body methods developed for the FQH problem to investigate FCI physics in moiré materials.

\emph{Adiabatic and Aharonov-Casher Hamiltonians ---}
We study the Hamiltonian
\begin{align}
    H_{\text{ad}}=H_{AC}+%U({\bm r})
    U=-\frac{1}{2m} \Pi_-\Pi_++U,
    \label{Adiabatic_Hamiltonian}
\end{align}
which, in a certain limit, approximates the continuum model of TMD homobilayers such as tMoTe$_2$ \cite{Yao_Skyrmions,morales-duran2024magic,AharonovCasher_TMD,crepel2024chirala,Fengcheng_variational}. In Eq.~\eqref{Adiabatic_Hamiltonian}, $\Pi_{\pm}= (p_x+A_x)\pm i(p_y+A_y)$ are kinetic-momentum operators with $p_\alpha = -i\partial_\alpha$ for $\alpha = x, y$. $U = U(\bmr)$ is a periodic potential (which we refer to as the residual potential), and $\bm A(\bmr) = (A_x(\bmr), A_y(\bmr))$ is a vector potential corresponding to a periodic magnetic field, $\nabla \times {\bm A}(\bmr)=B_0+B({\bm r})$,
which we have split into a spatially uniform part with one flux quantum per unit cell, $B_0=-1/\ell^2$, with $\ell$ the magnetic length, and a zero-average periodic piece $B(\bmr)$. We assume $B({\bm r})$ and $U({\bm r})$ have the same periodicity, with Fourier expansions
\begin{align}
    B({\bm r})=\sum_{{\bm G}}B_{\bm G}\,e^{i {\bm G}\cdot {\bm r}}, \quad\text{and}\quad U({\bm r})=\sum_{{\bm G}}U_{\bm G}\,e^{i {\bm G}\cdot {\bm r}},
    \label{PeriodicB}
\end{align}
where ${\bm G}$ are their shared reciprocal lattice vectors. See the Supplemental Material (SM) \cite{Supplemental} for further details on the Hamiltonian. 

Following \cite{AharonovCasher_TMD} we refer to Eq.~\eqref{Adiabatic_Hamiltonian} as the \textit{adiabatic Hamiltonian} and to its first term, $H_{AC}$, as the \textit{Aharonov-Casher Hamiltonian}, which is known to have a manifold of zero-energy states for arbitrary magnetic field \cite{aharonov1979ground, dubrovin1980ground}. When the magnetic field is periodic, the resulting zero-energy manifold can be understood as a quasi-Bloch band, which we will refer to as an AC band \cite{AharonovCasher_TMD,crepel2024chirala}. These flat bands are known to satisfy the trace condition, namely that their quantum metric $g^{ab}_{\bm k}$ and Berry curvature $\Omega_{\bm k}$ satisfy $\text{tr} g^{ab}_{\bm k}= \,|\Omega_{\bm k}|$, guaranteeing an exact mapping to the LLL \cite{roy2014band, jackson2015geometric}. As a consequence, AC bands can support Laughlin-like ground states whose analytic wavefunctions are related to those of the FQH effect \cite{wang2021exact,ledwith2020fractional,ledwith2023vortexabilitya}. 

The wave functions of the AC band can be written as
\begin{align}
    \psi^{AC}({\bm r})=f(z)\,e^{-\frac{|z|^2}{4\ell^2}+\chi({\bm r})}=\psi^{LLL}({\bm r})\,e^{\chi({\bm r})},
    \label{AC_wavefunction}
\end{align}
where $z=x+iy$, $f$ is a holomorphic function, $\chi({\bm r})$ is a real periodic function related to the magnetic field by $\nabla^2\,\chi({\bm r})=B({\bm r})$ and $\psi^{LLL}({\bm r})$ is a lowest Landau level wave function.
Eq. \eqref{AC_wavefunction} defines a mapping between LLL wave functions and AC wave functions, from which we can define a magnetic quasi-Bloch basis for the AC band $\ket{\psi_{\bm k}^{AC}}=e^{\chi({\bm r})} \,\ket{\psi_{\bm k}^{LLL}}$, in terms of the quasi-Bloch states of the LLL, $\ket{\psi_{\bm k}^{LLL}}$ \cite{ferrari1990twodimensional, ferrari1995wannier, haldane2018modularinvariant,wang2019lattice}. The Berry curvature of the AC band is given by \cite{wang2021exact,AharonovCasher_TMD}
\begin{align}
    \Omega_{\bm k}&=\frac{2\pi}{A_{\text{BZ}}}+\frac{1}{2}{\bm \nabla}_{\bm k}^2\ln \braket{\psi_{\bm k}^{AC}|\psi_{\bm k}^{AC}}\label{Eq:Berry_Curvature}\\
    &=\Omega_0+\frac{1}{2}{\bm \nabla}_{\bm k}^2\ln\left(\sum_{\bm G}\Phi_{\bm G} \braket{\psi^{\text{LLL}}_{\bm k}|e^{i \,{\bm G}\cdot\hat{{\bm r}}}|\psi^{\text{LLL}}_{\bm k}}\right),\nonumber
\end{align}
where $A_{\text{BZ}}$ is the area of the Brillouin zone and $\Phi_{\bm G}$ are the Fourier coefficients of the function $e^{2\chi (\bm r)}$. 
Eq.~\eqref{Eq:Berry_Curvature} shows that the Berry curvature distribution of the AC band can be controlled by tuning $B({\bm r})$, while the %quantum 
geometry is known to remain ideal \cite{wang2021exact}. When $U({\bm r})\neq 0$ in Eq. \eqref{Adiabatic_Hamiltonian}, the topmost band is no longer ideal, and we will refer to it as an adiabatic band. However, we expect the FQH-like ground state of the AC Hamiltonian to survive for a range of strengths of the residual potential.

\emph{Many-body calculations ---}
We fractionally fill the highest energy band of the Hamiltonian in Eq.~\eqref{Adiabatic_Hamiltonian} with spin-less holes and add band-projected un-screened Coulomb interactions, $V(\bmr) = e^2/\varepsilon|\bmr|$, with $\varepsilon$ the dielectric constant determining the interaction strength. The many-body Hamiltonian is
\begin{align}
    H=\sum_{{\bm k}} \epsilon_{{\bm k}}~ c^{\dagger}_{{\bm k}} c_{{\bm k}}+\frac{1}{2}\sum_{\substack{{\bm k}_i,{\bm k}_j\\{\bm k}_k,{\bm k}_l}} V_{ijkl} c^{\dagger}_{{\bm k}_i}c^{\dagger}_{{\bm k}_j}c_{{\bm k}_l} c_{{\bm k}_k},
\end{align}
where $\epsilon_{\bm k}$ are single-particle energies, $c^{\dagger}_{{\bm k}} (c_{{\bm k}})$ creates (destroys) a spin-less hole with momentum ${\bm k}$ and $V_{ijkl}$ is a two-particle matrix element. We then obtain the many-body spectrum by performing ED. The energy scales determining the spectrum are the interaction strength $U_{int}=e^2/\varepsilon\ell$, the cyclotron frequency $\omega_c=1/m\ell^2$ and, if present, the strength of the residual potential $U({\bm r})$.
%for the case of adiabatic bands. %%JC: I rewrote the previous sentence to remove "adiabatic bands" since the reader may not be familiar; I also added $U(r)$$ explicitly to remind what the residual potential refers to. 
See the SM \cite{Supplemental} for details on our numerics.

For simplicity we assume both functions in Eq. \eqref{PeriodicB} have six-fold symmetry and truncate them to the first shell of reciprocal lattice vectors, with magnitude $|{\bm G}|=G_0$. In this approximation, the first-shell harmonics, denoted $B_1$ and $U_1$, serve as tuning parameters that control the Berry curvature distribution and deviations from ideal geometry of the topmost band of Eq.~\eqref{Adiabatic_Hamiltonian}. In the SM \cite{Supplemental} we justify this approximation by 
showing that the higher harmonics $B_2,\, U_2, \dots$ only weakly affect the many-body spectra.
In the following, $B_1$ has units of $1/\ell^2$ and $U_1$ has units of $\omega_c$. For calculations presented in the main text, we focus on hole-filling $\nu=1/3$; in the SM \cite{Supplemental} we also consider hole-filling $\nu=2/3$. We observe particle-hole asymmetry with respect to half-filling for both $B_1\neq 0$ and $U_1\neq 0$, which is expected due to the non-trivial quantum geometry of the topmost band \cite{BergholtzMoessner,abouelkomsan2023quantum}.

\emph{Effects of inhomogeneous Berry curvature on ideal band FCIs ---}
We first study the many-body spectrum of AC bands, which are obtained from the Hamiltonian in Eq. \eqref{Adiabatic_Hamiltonian} with $U({\bm r})=0$. Despite the AC band being ideal, an exact zero-energy Laughlin-like ground state at $\nu=1/3$ is not required, as the interaction we chose contains terms beyond the first Haldane pseudopotential. While we observe an FCI ground state for $B(\bm{r}) = 0$, strong Berry curvature inhomogeneity drives a transition out of that state.
Previous studies of this transition simultaneously included deviations from the ideal band condition \cite{KaiSun_Hofstadter,Repellin_Senthil,Lauchli_FCICDW, LauchliIdeal,Shavit_QGStability}. 
The unique aspect of our work is that we have isolated the effect of Berry curvature fluctuations from deviations from idealness on the nature of the ground state. We achieve this by tuning $B_1$, which determines the Berry curvature of the ideal AC band via Eq. \eqref{Eq:Berry_Curvature}. 

\begin{figure}
    \centering   \includegraphics[width=0.48\textwidth]{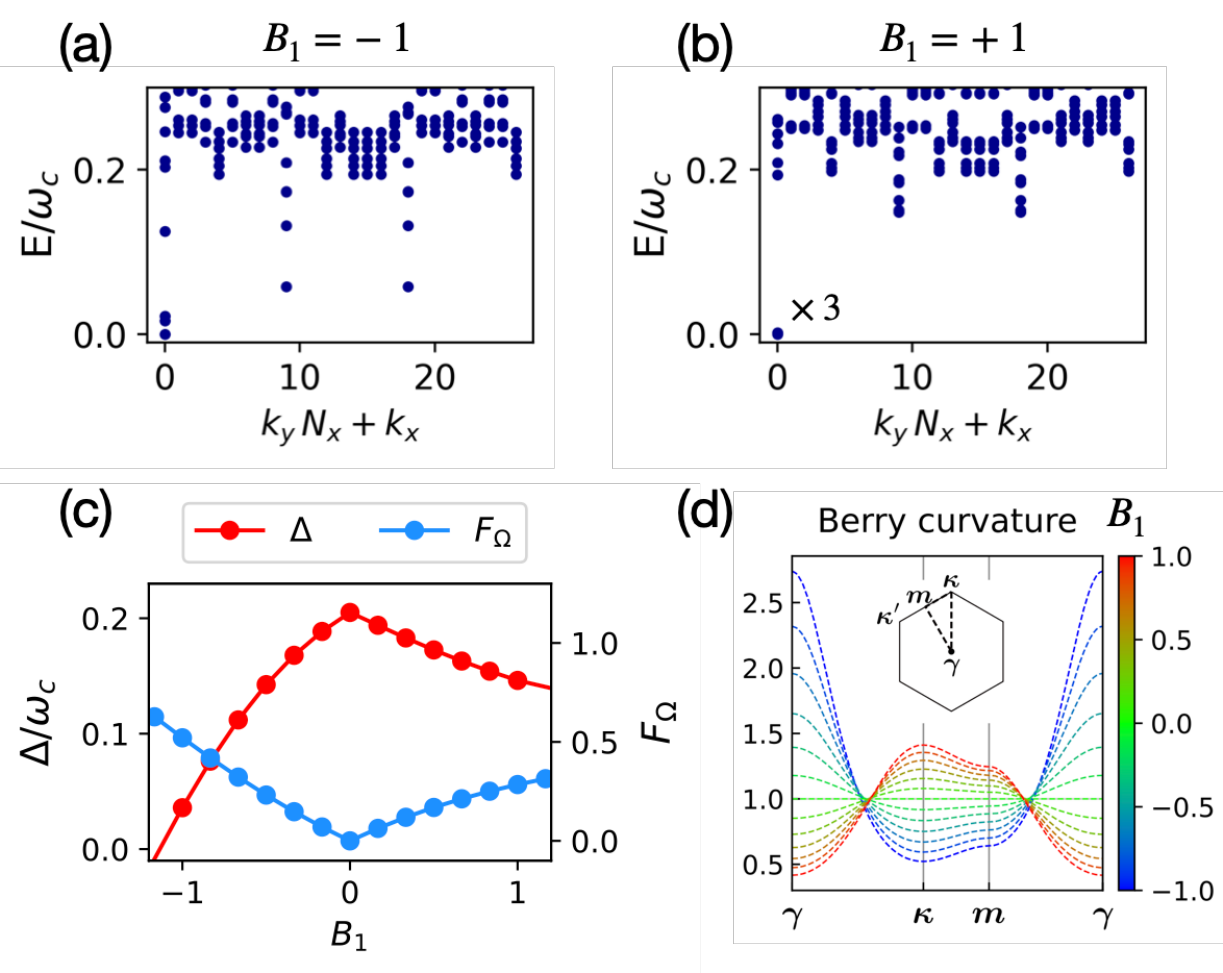}
    \caption{(a)-(b) Many-body spectra at filling $\nu=1/3$ for an AC band with $B_1=-1$ and $B_1=+1$, respectively. (c) Dependence of the magnetoroton gap $\Delta$ at filling $\nu=1/3$ (red) and Berry curvature standard deviation $F_{\Omega}$ (see Eq. \eqref{eq:BC_Stdev}) of the AC band (blue) as a function of $B_1$. (d) Berry curvature distribution of the AC band as a function of $B_1$. Calculations were performed for systems with 27 unit cells.}
    \label{fig:Berry_Fluctuations}
\end{figure}

Figs. \ref{fig:Berry_Fluctuations}(a)-(b) show the many-body spectrum at filling $\nu=1/3$ for AC bands with $B_1=-1$ and $B_1=+1$, respectively. For both calculations the ground state remains in the same universality class, with three quasi-degenerate ground states in the same momentum sector, but the Berry curvature distribution significantly affects the low-energy excitation spectrum. Specifically, the gap to the lowest neutral excitation at ${\bm \kappa/\bm \kappa^{\prime}}$, which we identify as a magnetoroton \cite{Repellin_MRM}, is much smaller when $B_1$ is negative. 
Fig. \ref{fig:Berry_Fluctuations}(c) shows this asymmetry explicitly by plotting the neutral gap, $\Delta$, as a function of $B_1$; see SM \cite{Supplemental} for the definition of $\Delta$.

We quantify Berry curvature inhomogeneity of the band through 
\begin{align}
    F_{\Omega}=\sqrt{\int_{\text{BZ}}\frac{d^2{\bm k}}{A_{\text{BZ}}}\left(\frac{\Omega_{\bm k}}{\Omega_0}-C\right)^2},
    \label{eq:BC_Stdev}
\end{align}
where $C$ is the Chern number and $\Omega_0$ the average Berry curvature. Fig. \ref{fig:Berry_Fluctuations}(c) shows $F_{\Omega}$ as a function of $B_1$, while Fig. \ref{fig:Berry_Fluctuations}(d) shows the Berry curvature distribution of the AC band across the Brillouin zone, as a function of $B_1$. The location of the Berry curvature peaks in Fig. \ref{fig:Berry_Fluctuations}(d) follows from a perturbative expansion of Eq. \eqref{Eq:Berry_Curvature} in $B_1$, detailed in the SM \cite{Supplemental},
\begin{align}
    \Omega_{\bm k}\approx\Omega_0-\frac{4\pi \ell^2\,B_1\,e^{-\frac{G_0^2\ell^2}{4}}}{\sqrt{3}G_0^2}\,\sum_{j=1}^6e^{i\,{\bm k}\cdot {\bm R}_j} + O(B_1^2),
    \label{eq:Berry_perturb}
\end{align}
here ${\bm R}_j$ are the six first-shell lattice vectors determining the periodicity of $B({\bm r})$. We see that for $B_1>0$, Eq. \eqref{eq:Berry_perturb} predicts that the Berry curvature has peaks at $\bm\kappa/\bm\kappa'$ while for $B_1<0$ it predicts a single peak at $\bm\gamma$. 
However, Fig. \ref{fig:Berry_Fluctuations}(d) also reveals an asymmetry in the magnitude of peaks/troughs beyond the linear expansion in Eq.~\eqref{eq:Berry_perturb}.

Fig.~\ref{fig:Berry_Fluctuations}(c) shows that the maximum magnetoroton gap $\Delta$ is attained for homogeneous Berry curvature, \textit{i.e.} $B_1=0$, and that fluctuations in Berry curvature lead to an approximately linear decrease in $\Delta$ where the slope depends on the sign of $B_1$. To understand this behavior, we use the single mode approximation to define low-energy neutral excitations on top of the AC ground state. As discussed in the SM \cite{Supplemental}, discrete translational symmetry couples states that differ by a reciprocal lattice vector, causing the degenerate states at the ${\bm \kappa}/{\bm \kappa^{\prime}}-$points to split linearly with $B_1$, 
while generically other states receive a correction of order $B_1^2$.
Consequently, for small $B_1$, the lowest energy excitation occurs at ${\bm \kappa}/{\bm \kappa^{\prime}}$.
%leading to the minimum excitation to happen at the ${\bm \kappa}/{\bm \kappa^{\prime}}-$points. 
In the limit $B_1\ll 1$, we apply degenerate perturbation theory to the magnetoroton states at ${\bm \kappa}/{\bm \kappa^{\prime}}$ and
obtain an expression for the AC magnetoroton gap in terms of the LLL gap, denoted $\Delta^{LLL}$:
\begin{align}
    \Delta({\bm \kappa})\approx\begin{cases}
        \Delta^{LLL}({\bm \kappa})-\alpha\, |B_1|, \quad\text{for } B_1<0.\\
        \Delta^{LLL}({\bm \kappa})-\beta\, |B_1|, \quad\text{ for } B_1>0.
    \end{cases}
    \label{eq:AC_magnetorotongap}
\end{align}
See the SM \cite{Supplemental} for details on our magnetoroton model for AC bands and expressions for the coefficients $\alpha$ and $\beta$, which satisfy $\alpha>\beta>0$, according to our numerics. Eq. \eqref{eq:AC_magnetorotongap} captures the asymmetric behavior of $\Delta$ with respect to $B_1=0$ observed in Fig. \ref{fig:Berry_Fluctuations}(c). For completeness, in the SM \cite{Supplemental} we present momentum occupation plots $n_{\bm k}$, which show a smaller occupation in regions of the Brillouin zone where the Berry curvature is peaked, in agreement with \cite{abouelkomsan2023quantum}. Ultimately the large inhomogeneity in $\Omega_{\bm k}$, which translates into a large $F_{\Omega}$, leads to a charge density wave (CDW) with $\Delta<0$ for $B_1<0$, as can be seen in Fig. \ref{fig:Berry_Fluctuations}(c). 

\emph{Effects of inhomogeneous Berry curvature on adiabatic band FCIs ---}
In contrast to the AC bands studied in the previous section, bands that describe the low-energy physics of two-dimensional materials generically have a non-vanishing dispersion and do not satisfy the trace condition. To recover that scenario, we restore the residual potential $U({\bm r})$ in the adiabatic Hamiltonian, Eq. \eqref{Adiabatic_Hamiltonian}, yielding a non-ideal dispersive topmost band with non-uniform Berry curvature. Tuning the parameter $U_1$ will simultaneously change the Berry curvature distribution, destroy the trace condition and introduce a finite band width. These effects together ultimately determine the interacting ground state.

\begin{figure}
    \centering   \includegraphics[width=0.45\textwidth]{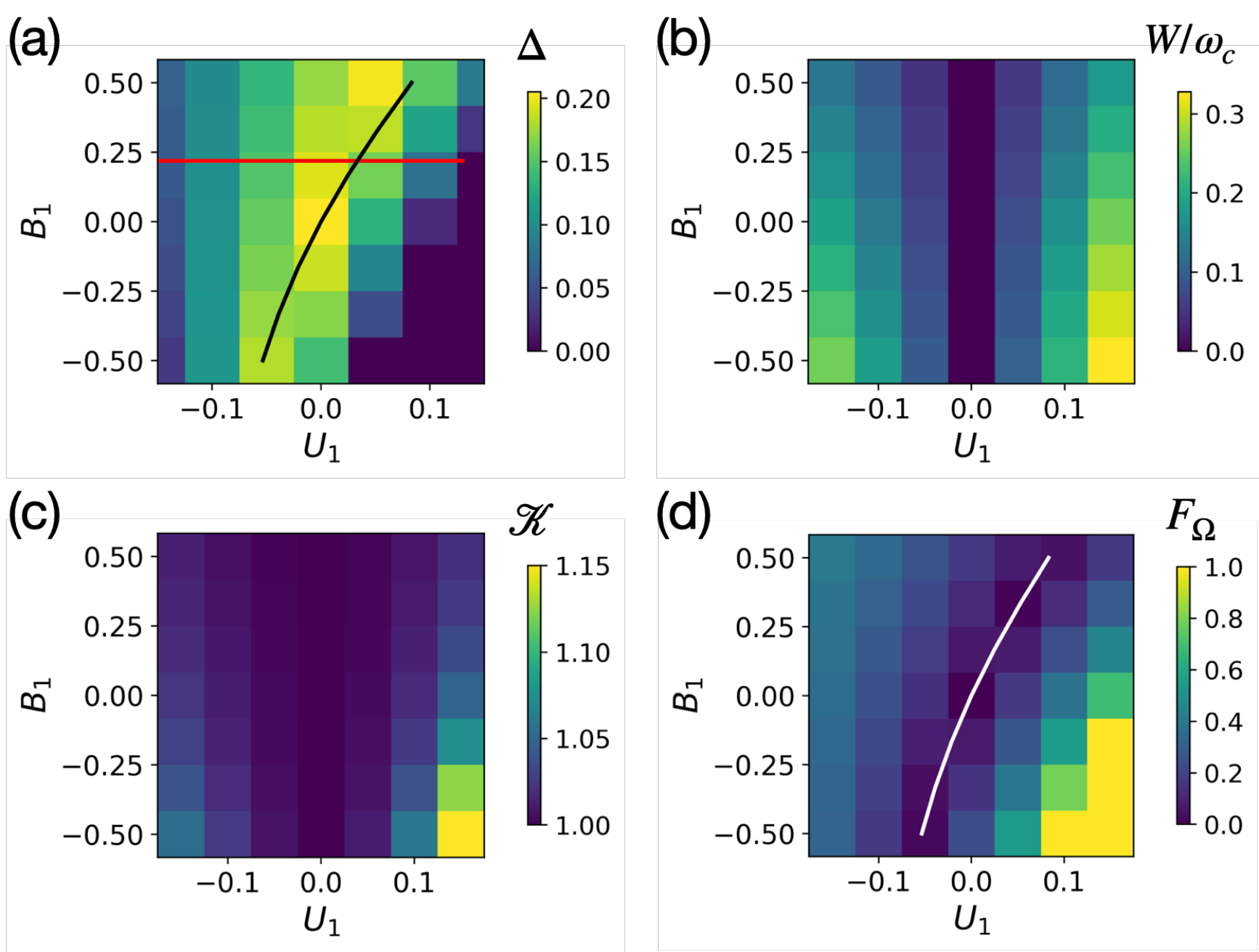}
    \caption{(a) Magnetoroton gap $\Delta$ as a function of $U_1$ and $B_1$; $\Delta\le 0$ indicates that the FCI is not the ground state. The red line indicates the parameters obtained from applying the adiabatic approximation to a continuum model of twisted MoTe$_2$ \cite{FCI_DiXiao} for twist angles $\theta\sim 2.5^{\circ}-5.0^{\circ}$. The black line traces the minimum value of $F_{\Omega}$. (b) Band width $W$, (c) quantum weight $\mathcal{K}$ (see main text) and (d) Berry curvature standard deviation for adiabatic bands, as a function of $U_1$ and $B_1$. The white line in (d) indicates the trajectory of the minimum value of $F_{\Omega}$ and is also plotted as a black solid line in (a). The optimal lines in (b)-(c) correspond to the axis $U_1=0$.}
    \label{fig:Trace_Deviations}
\end{figure}

Fig. \ref{fig:Trace_Deviations}(a) shows the evolution of $\Delta$ as a function of $B_1$ and $U_1$. The region of FCI stability is determined by $\Delta>0$. Figs. \ref{fig:Trace_Deviations}(b)-(d) show the band width $W$, quantum weight $\mathcal{K}$ and Berry curvature inhomogeneity $F_{\Omega}$, respectively, for adiabatic bands as a function of $B_1$ and $U_1$. The quantum weight is defined as the Brillouin zone average of ${\rm tr}g_{\bm k}$ divided by $2\pi$ \cite{QWeight1,QWeight2,QWeight3}, which equals the Chern number $C$ for ideal bands and thus measures the violation of trace condition. Both $W$ and $\mathcal{K}$ are optimized at $U_1=0$ and are larger for $U_1>0$ than for $U_1<0$, regardless of the sign of $B_1$. Deviations from the optimal value of these quantities contribute to the reduced stability of the FCI. The data reveals a surprising observation: for fixed $B_1 > 0$, the largest many-body gap occurs not for the ideal band $(U_1 = 0)$, but for a non-ideal band with finite $U_1>0$.
Similarly, for fixed $B_1 < 0$, the largest many-body gap occurs for a non-ideal band with $U_1<0$. This results in a larger region of FCI stability for bands with $B_1> 0$. Comparing Figs. 2(a)-(d) reveals that the maximum many-body gap approximately coincides with the minimum in $F_\Omega$, indicating that the Berry curvature variation may be more important than satisfying the ideal band condition in this nearly-ideal parameter regime. Our numerical results are consistent with a perturbative calculation of the many-body gap using the SMA, detailed in the SM~\cite{Supplemental}. 

The behavior of $F_{\Omega}$ for adiabatic bands can be understood as follows: The Berry curvature induced by $U_1>0$ ($U_1<0$) and $B_1>0$ ($B_1<0$) is peaked in opposite locations in the Brillouin zone, so that the adiabatic Hamiltonian with finite $U_1, B_1$ has smaller Berry curvature fluctuations than when either term separately vanishes, in the SM \cite{Supplemental} we present plots for the Berry curvatures of the adiabatic bands to illustrate this behavior. We note that in our analysis it is crucial that $U_1$ is small (namely $|U_1|\ll \omega_c$ and smaller than the many-body gap at $U_1=0$), in contrast to previously studied topological band models where flattening the Berry curvature can destabilize the FCI phase \cite{VarjasFlatBC} but the trace condition is strongly violated.

\emph{Application to moiré materials ---} 
The adiabatic Hamiltonian studied here gives insight into the stability of FCIs in twisted homobilayer TMDs. Specifically, the continuum model for the latter can be mapped to Eq. \eqref{Adiabatic_Hamiltonian} by replacing the layer degree of freedom by a periodic magnetic field and a periodic potential, and discarding all but the lowest harmonics of $B({\bm r})$ and $U({\bm r})$. We apply the adiabatic approximation described in Refs. \cite{morales-duran2024magic,AharonovCasher_TMD} to a particular model of twisted MoTe$_2$ (tMoTe$_2$) \cite{FCI_DiXiao}, yielding $B_1=0.22$ [$1/\ell^2$], which is independent of twist angle, and $U_1=-7.36+0.57\,(\theta[\text{deg}])^2$ in units of meV. By converting to units of $\omega_c$, we obtain that for the experimentally relevant range $\theta \sim 2.5^{\circ}-5.0^{\circ}$, the value of $U_1$ goes from $U_1=-0.27$ [$\omega_c$] to $U_1=0.13$ [$\omega_c$]. The obtained values of the adiabatic parameters are indicated by a solid red line in Fig \ref{fig:Trace_Deviations}(a), justifying the parameter ranges that we have chosen for our study. In the SM \cite{Supplemental} we present a comparison between the continuum model and its adiabatic approximation, showing that the adiabatic Hamiltonian, Eq. \eqref{Adiabatic_Hamiltonian}, reproduces the main features of the continuum moiré Hamiltonian, both at the single-particle and many-body levels. The result that the many-body gap $\Delta$ is non-zero across nearly the entire red line in Fig.~\ref{fig:Trace_Deviations}(a) provides a possible explanation for why the FCI phase in semiconductor moiré systems remains stable over a large range of twist angles, as numerically \cite{Princeton_ED_FCI,Reddy_GlobalPhaseDiagram,reddy2023fractional,morales-duran2023pressureenhanced,FCI_DiXiao,li2021spontaneous} and experimentally \cite{cai2023signatures, zeng2023thermodynamic, park2023observation, xu2023observation} observed;
and predicts that for angles above the magic angle, where the topmost continuum band is closest to flat and ideal, the FCI ground state will be more robust.

\emph{Discussion ---} We have numerically and analytically investigated the effect of Berry curvature fluctuations on the many-body gap of FCIs in ideal and near-ideal bands, in the presence of a realistic Coulomb interaction. We derived perturbative expressions for the FCI magnetoroton gaps in terms of the LLL magnetoroton gap, which explain the asymmetry of the neutral FCI gap with respect to the momentum-space position of the Berry curvature maxima, for both AC and adiabatic bands. This powerful analytical result sheds light on the properties of Berry curvature that favor the stability of FCI ground states in nearly-ideal bands and is complementary to other approaches to address FCI excitations \cite{Repellin_MRM,KaiSun_MRM1,KaiSun_MRM2,ZiYang_MRM,ChongWang_MRM_ED,MRM_DMRG1,Hyperdeterminants,Fengcheng_MoireAssisted,Toby_Structurefactor,Debanjan_StructureFactor,LiangFu_MRM}. By exploiting the adiabatic approximation, we have used our results to explain the large region of stability for topologically ordered phases in twisted TMDs, where the band geometry is no longer ideal. This connection establishes a framework which provides analytical insights into the role of band geometry on FCIs in real materials.
 
We focused on spinless fermions, for which the FCI is destabilized by neutral magnetoroton modes. When spin or orbital degrees of freedom are taken into account, other candidate phases can become competitive \cite{Repellin_Senthil,Regnault_FTI}. In the SM \cite{Supplemental} we present results showing that the valley-flip excitations have higher energies than the lowest valley-polarized excitations, which justifies the restriction to a single valley in this work. Our band-projected results yield finite and comparable magnetoroton gaps for $\nu=1/3$ and $\nu=2/3$ (See SM \cite{Supplemental}.) This is in contrast to experiments in tMoTe$_2$, where the FCI state is robust at $\nu=2/3$ and absent at $\nu=1/3$, pointing to the importance of remote band mixing in realistic systems, as discussed by other authors \cite{abouelkomsan2024band, Bernevig_bandmixing}. We leave the analysis of remote band effects in our model for future work.\\

{\it Acknowledgements:} The authors are grateful to Allan H. MacDonald for related collaborations and inspiring discussions. NMD thanks Valerio Peri, Pawel Potasz, Aidan Reddy, Nicolas Regnault, Cécile Repellin and Jie Wang for discussions. JS thanks Xiaodong Hu and Xiaoyang Shen for helpful interactions. This research was supported in part by grant NSF PHY-2309135 to the Kavli Institute for Theoretical Physics (KITP). JS acknowledges support by the US Department of Energy, Office of Science, Basic Energy Sciences, Materials Sciences and Engineering Division. JC acknowledges support from the Air Force Office of Scientific Research under Grant No. FA9550-24-1-0222 and from the Alfred P. Sloan Foundation through a Sloan Research Fellowship. We acknowledge computational resources provided by the Texas Advanced Computing Center (TACC). The Flatiron Institute is a division of the Simons Foundation.

{\it Author contributions:} NMD and JS designed the project and JC supervised it. JS calculated the non-interacting band structures and provided the basis wave functions for exact diagonalization; NMD and JC developed the magnetoroton theory; NMD performed the ED calculations and wrote the paper with input from all authors.

\bibliographystyle{apsrev4-1}
\bibliography{bibliography}

\clearpage

\appendix
\onecolumngrid
\section*{Supplemental material for ``Effects of Berry curvature on ideal band magnetorotons"}

\section{Derivation of Aharonov-Casher Hamiltonian}

The original Aharonov-Casher model describes a 2D system of spin-1/2 electrons in a periodic magnetic field \cite{aharonov1979ground,dubrovin1980ground}, with the full single-particle Hamiltonian
\begin{equation}
    H = - \left( \frac{1}{2m} {\bm\Pi}^2 + \frac{B}{2m}\sigma^z \right).
\end{equation}
Here ${\bm\Pi} = {\bm p} + {\bm A} = (p_x + A_x,\, p_y + A_y)$ is the kinetic momentum vector operator, $B = B_0 + B({\bm r}) = \nabla \times {\bm A}$ is the total out-of-plane magnetic field and $\sigma^z$ is the spin Pauli matrix. (The overall minus sign is chosen to align with our convention that connects to the valence bands of TMDs.) Since $\sigma^z$ is conserved, 
the spins decouple:
%the system can be decoupled in to spin-up ($\sigma^z = 1$) and spin-down ($\sigma^z = -1$) sectors, the former one identical to our $H_{AC}$ in Eq. (\ref{Adiabatic_Hamiltonian}):
%
\begin{equation}
    H_{\sigma^z = \pm 1} = -\frac{1}{2m} \Pi_\mp \Pi_\pm,
\end{equation}
where the spin-up ($\sigma^z = 1$) sector corresponds to $H_{AC}$ in Eq. (\ref{Adiabatic_Hamiltonian}) in the main text.
Both $\Pi_+ = p_x + A_x + i(p_y + A_y) = -2i\partial_{z^*} + A({\bm r})$ and $\Pi_- = p_x + A_x - i(p_y + A_y) = -2i\partial_z + A^*({\bm r})$ (here $z = x + iy$) have a family of zero eigenfunctions, denoted $\phi_\pm({\bm r})$ (i.e. $\Pi_\pm \phi_\pm(\bm r) = 0$), that take the general forms
\begin{equation}
    \phi_+(\bm r) = f(z) e^{\alpha(\bm r)}, \quad \phi_-(\bm r) = f'(z^*) e^{-\alpha^*(\bm r)},
\end{equation}
where $f$ and $f'$ are arbitrary holomorphic functions and $\alpha$ satisfies $\partial_{z^*}\alpha = -iA(\bm r)/2$. In our convention, $A(\bm r) = iB_0z/2 + A'(\bm r) = -iz/2\ell^2 + A'(\bm r)$ contains a $z$-linear part $-iz/2\ell^2$ and a periodic part $A'(\bm r)$, which means that $\alpha$ contains a quadratic part $-z^*z/4\ell^2$, a periodic part $\chi(\bm r)$ satisfying $\partial_{z^*}\chi = -iA'(\bm r)/2$, and an arbitrary $z$-holomorphic part that can be absorbed into $f$ or $f'$. Due to the quadratic part, only $\phi_+(\bm r)$ is normalizable and gives physical wave functions. Hence, only in the $\sigma^z = +1$ sector does the system have zero-energy states. In this work, the model is limited to $\sigma^z = 1$ so that the filling holes are effectively spinless.\\\\

\subsection{Single-particle and many-body results of AC, adiabatic and twisted TMD models}

\begin{figure}
    \centering   \includegraphics[width=0.45\textwidth]{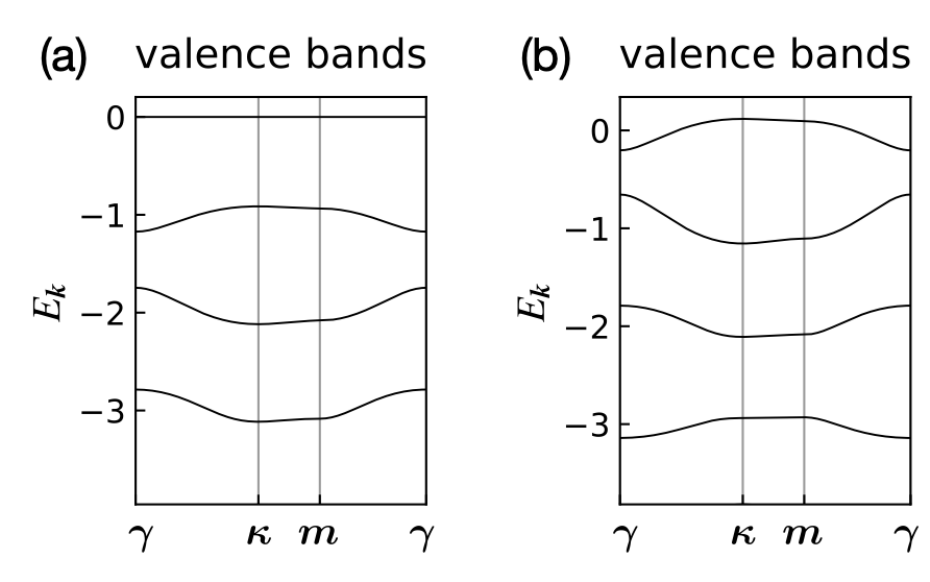}
    \caption{Band structures including the top 4 bands of (a) AC model with $B_1 = 1/6$, and (b) adiabatic model with $B_1 = 1/6$, $U_1 = 0.2$. $E_{\bm k}$ is in units of $\omega_c$.}
    \label{fig:AC_Adiabatic_Bandstructures}
\end{figure}

\begin{figure}
    \centering   \includegraphics[width=0.75\textwidth]{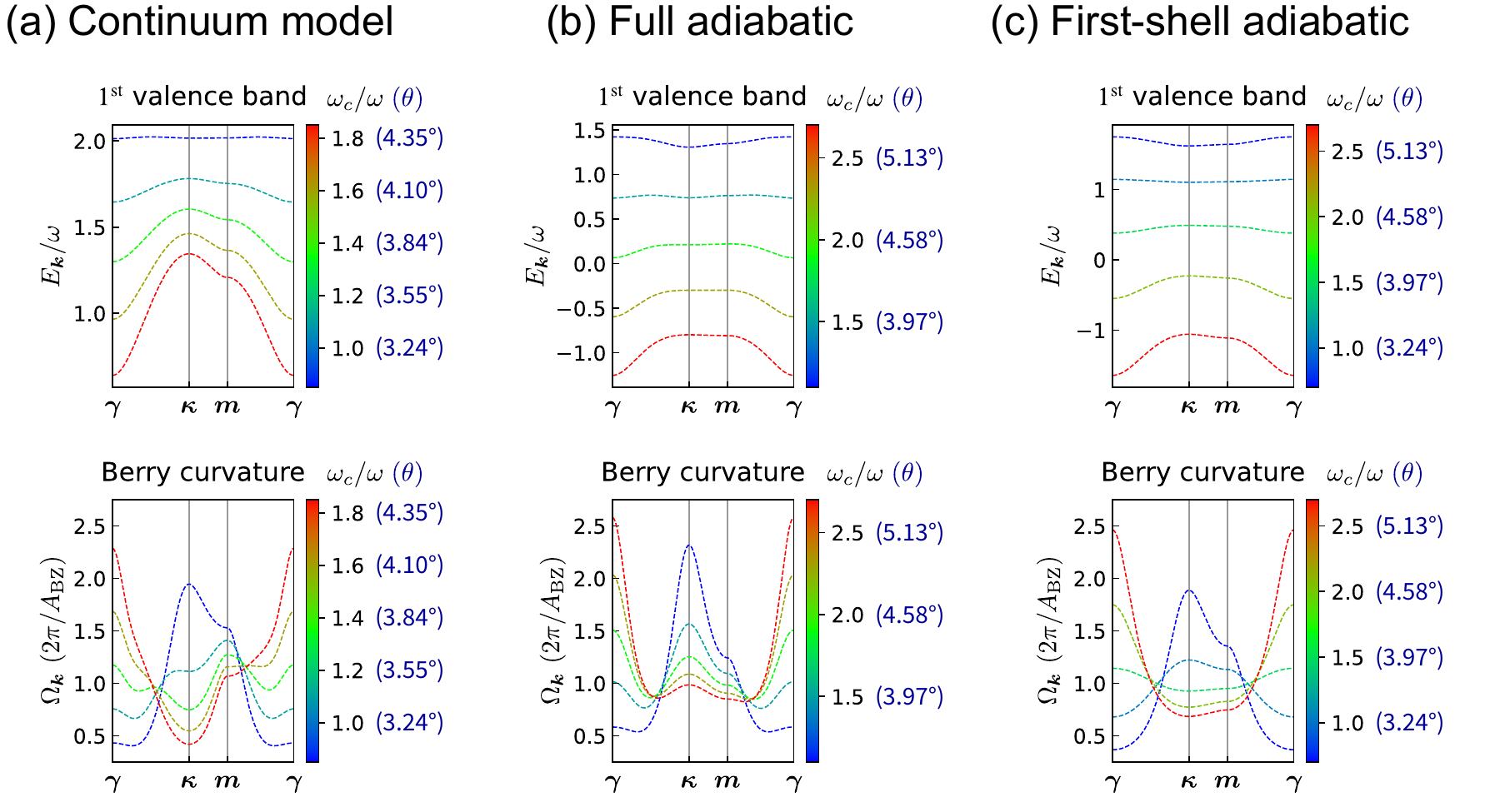}
    \caption{Band structure and Berry curvature distribution for (a) a continuum model for MoTe$_2$ \cite{FCI_DiXiao}, (b) the adiabatic approximation \cite{morales-duran2024magic,AharonovCasher_TMD} to the continuum model including 25 harmonics in the periodic functions $B(\bm r)$ and $U(\bm r)$, and (c) the adiabatic approximation where both $B(\bm r)$ and $U(\bm r)$ are truncated to only the first harmonic, which corresponds to the model studied in the present work. We have taken $\omega=23.8$ meV \cite{FCI_DiXiao} as the unit of energy and present plots as a function of twist angle. The relation between the cyclotron gap and the twist angle is $\hbar\omega_c=(4\pi^2\hbar^2/\sqrt{3}ma_0^2)\theta^2$ \cite{morales-duran2024magic,AharonovCasher_TMD}, with $a_0$ the MoTe$_2$ monolayer lattice constant and $m$ the effective mass.}
    \label{fig:Continuum_vs_Adiabatic_singleparticle}
\end{figure}

\begin{figure}
    \centering   \includegraphics[width=0.85\textwidth]{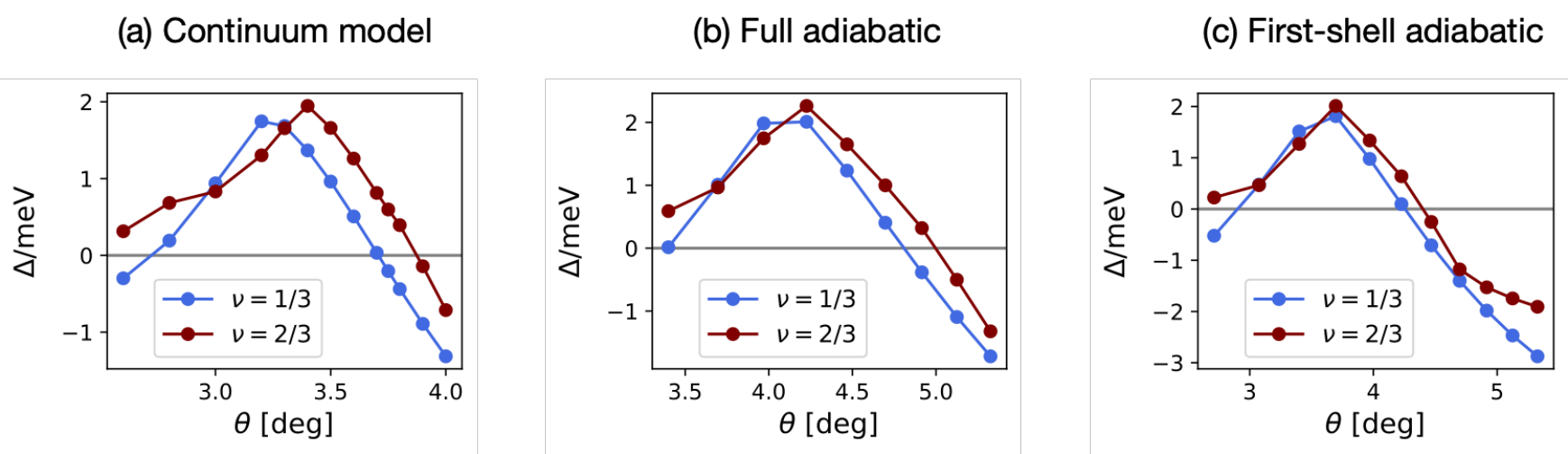}
    \caption{Many-body gap obtained from ED calculations for (a) the continuum model of MoTe$_2$ \cite{FCI_DiXiao}, (b) the adiabatic approximation to the same continuum model including 25 harmonics in the periodic functions $B({\bm r})$ and $U({\bm r})$, and (c) the adiabatic approximation with $B({\bm r})$ and $U({\bm r})$ truncated to the first harmonic. Both filling fractions $\nu=1/3$ and $\nu=2/3$ are shown and we have used $\varepsilon=20$ for all calculations.}
    \label{fig:Continuum_vs_Adiabatic_manybody}
\end{figure}

Two families of models are involved in this work. The first one is the adiabatic model given by Eq. (\ref{Adiabatic_Hamiltonian}) in the main text, in which $B_1$ and $U_1$ are independently scanned. Fig. \ref{fig:AC_Adiabatic_Bandstructures} shows %a
typical band structures of this model both in and away from the AC limit. While the topmost valence band of the AC Hamiltonian is flat, adding the periodic potential introduces a finite band width and violates the trace condition. \\\\
The second family of models that we study are based on the continuum model of tMoTe$_2$ \cite{FCI_DiXiao}, we adopt different levels of approximations while the only scanning parameter is the twist angle $\theta$. Fig. \ref{fig:Continuum_vs_Adiabatic_singleparticle} (a) shows the single-particle dispersion and Berry curvature of the first valence band as functions of $\theta$, computed directly from the original continuum model; Fig. \ref{fig:Continuum_vs_Adiabatic_singleparticle} (b) shows the same quantities from the model obtained by applying the adiabatic approximation \cite{morales-duran2024magic, AharonovCasher_TMD} to the continuum model, which yields a Hamiltonian of the same form as Eq. (\ref{Adiabatic_Hamiltonian}) in the main text, except that the shape of $B(\bmr)$ is now fixed by the tMoTe$_2$ parameters (and $U(\bmr)$ is the sum of two pieces with fixed individual shapes and $\theta$-tuned relative scale \cite{AharonovCasher_TMD}); Fig. \ref{fig:Continuum_vs_Adiabatic_singleparticle} (c) shows these quantities from the same adiabatic approximation followed by a further step of approximation which truncates $B(\bmr)$ and $U(\bmr)$ to the first shells $B_1$ and $U_1$, directly connecting to our first family of AC and adiabatic models (again, here $B_1 = 0.22\,[1/\ell^2]$ is fixed by the tMoTe$_2$ parameters, and only $U_1$ is tunable by $\theta$. See main text.) We find that the adiabatic models, whether or not truncated to the first shell, well reproduce the trends in band structure and Berry curvature variations present in the continuum model. For completeness, in Fig. \ref{fig:Continuum_vs_Adiabatic_manybody} we also compare these three models at the many-body level (see the following section for details on how the many-body results are obtained) by showing the evolution of the many-body gap with twist angle, for the relevant filling fractions $\nu=1/3$ and $\nu=2/3$. 
The adiabatic approximation reproduces the main features of the continuum model, especially at twist angles $\theta \lesssim 4^{\circ}$, as expected \cite{morales-duran2024magic,AharonovCasher_TMD}.


\section{Many-body calculation methods}
A natural basis in which to write the eigenstates of the Hamiltonian given by Eq. \eqref{Adiabatic_Hamiltonian} in the main text is that of the Landau levels of the magnetic field $B_0$,
\begin{align}
    \psi_{n,{\bm k}}({\bm r})=\sum_{m}c_{m,{\bm k}}^n \psi_{\bm k}^{mLL}({\bm r}),
    \label{eq:AC+U_eigenstate}
\end{align}
where $n$ is a band index, $m$ a Landau level index, and $\bm k$ is the quasi-Bloch momentum that characterizes the magnetic translational eigenstates. The coefficients $c_{m,\bm k}^n$ can be obtained by a diagonalization scheme developed in Ref. \cite{AharonovCasher_TMD}. In order to obtain the many-body spectrum of Eq. \eqref{Adiabatic_Hamiltonian} we require to compute the two-particle interaction matrix elements between the eigenstates defined by Eq. (\ref{eq:AC+U_eigenstate}). The twisted Bloch periodicity of these single-particle eigenstates due to the magnetic flux is
\begin{equation}
    \psi_{n,\bm k}(\bmr + \bmR) = \eta_\bmR e^{i \left( 
\bmk \cdot \bmR + \frac{\bmR \times \bmr}{2} \right)} \psi_{n, \bmk}(\bmr),
    \label{eq:twistedBloch}
\end{equation}
(here $\bmR$ is arbitrary lattice vector and $\eta_\bmR$ is the parity signature, \textit{i.e.}, 1 if $\bmR/2$ is in the Bravais lattice and $-1$ otherwise), which complicates the direct evaluation of interaction matrix elements. For that reason we introduce an auxiliary spinor field $u_s(\bmr)$ which cancels out the magnetic twist to give a perfect Bloch state:
\begin{equation}
    \phi_{n, \bmk s}(\bmr) = \psi_{n,\bmk}(\bmr) u_s(\bmr) = \sum_{\bmG} z_{n,\bmk,\bmG,s} e^{i (\bmk + \bmG) \cdot \bmr},
    \label{eq:eigenstate_mapped}
\end{equation}
where $s = +, -$ is a pseudospin index and $\bmG$ is a reciprocal lattice vector. We emphasize here that the pseudospin index $s$ is purely auxiliary and does not need to be associated with any physical flavor. The choice of $u_s(\bmr)$ has large freedom: as long as it has the proper twisted periodicity and normalization $|{\bm u}(\bmr)|^2=1$, the determination of interaction matrix elements can be carried out in the Bloch basis defined in Eq. (\ref{eq:eigenstate_mapped}), which lives in the Hilbert space of the continuum model of twisted homobilayer TMDs \cite{wu2019topological}. This mapping also facilitates comparison with many-body numerical calculations using realistic continuum models of WSe$_2$ or MoTe$_2$ with parameters obtained from DFT \cite{Fengcheng_variational}. Importantly, we note that the effective interlayer and intralayer interactions in our ``fictitious bilayer'' have to be both identical to the original interaction, thereby ensuring that the interaction matrix elements do not depend on the details of $u_s(\bmr)$.\\\\
We now describe the specific choice of $u_s(\bmr)$ for all our many-body calculations, which derives from the adiabatic approximation of twisted homobilayer TMDs \cite{morales-duran2024magic, AharonovCasher_TMD}. We locally diagonalize the layer-resolved moir\'e potential:
\begin{subequations}
\begin{equation}
    \Delta(\bmr) = \Delta_0(\bmr) + {\bm\sigma} \cdot {\bm\Delta}(\bmr),
\end{equation}
\begin{equation}
    \Delta(\bmr) \tilde{\bmu}(\bmr) =
    \left( \Delta_0(\bmr) + |{\bm\Delta}(\bmr)| \right) {\tilde{\bmu}}(\bmr),
\end{equation}
\end{subequations}
where $\bm\sigma = (\sigma^x, \sigma^y, \sigma^z)$ is the layer-pseudospin Pauli matrix vector and we define the functions
\begin{subequations}
\begin{equation}
    \Delta_0(\bmr) \pm \Delta_z(\bmr) = 2V \sum_{j=0}^2 \cos (\bmG_{2j} \cdot \bmr \mp \psi),
\end{equation}
\begin{equation}
    \Delta_x(\bmr) \pm i\Delta_y(\bmr) = w\sum_{j=0}^2 e^{\pm i {\bm q}_j \cdot \bmr},
\end{equation}
\end{subequations}
with ${\bm q}_j = G_0/\sqrt{3}\,( \sin(2j\pi/3), \, -\cos(2j\pi/3) )$ and $\bmG_j = G_0 ( \cos(j\pi/3), \, \sin(j\pi/3) )$ where $G_0$ is the magnitude of the first-shell reciprocal lattice vectors, and $(w, V, \psi)$ are three material-dependent phenomenological parameters. When $w, V > 0$ and $0 < \psi < \pi$, the vector field $\bm\Delta(\bmr)$ has a topologically non-trivial real-space texture leading to a unit real-space Chern number of $\bmu(\bmr)$.\\\\
We numerically fix the U(1) gauge of $\Tilde{\bmu}(\bmr)$ so that the real-space Berry connection $\bmA_{ad}(\bmr)$ is a specific solution to $\nabla \times \bmA_{ad} = B_{ad}(\bmr) = {\bm\Delta} \cdot (\partial_x \bm\Delta \times \partial_y \bm\Delta) / 2|\bm\Delta|^3$:
\begin{equation}
    A_{ad}(\bmr) = A_{ad}^x(\bmr) + iA_{ad}^y(\bmr) = -i \begin{pmatrix}
        \Tilde{u}_+^*(\bmr) & \Tilde{u}_-^*(\bmr)
    \end{pmatrix} \begin{pmatrix}
        (\partial_x + i\partial_y) \Tilde{u}_+(\bmr) \\
        (\partial_x + i\partial_y) \Tilde{u}_-(\bmr)
    \end{pmatrix} = -\frac{i(x+iy)}{2} + \sum_{\bmG\ne 0} \frac{B_{ad,\bmG}}{G_x - iG_y} e^{i\bmG\cdot\bmr},
    \label{eq:A_ad}
\end{equation}
which satisfies the Coulomb gauge $\nabla\cdot\bmA = 0$. We note here that the real-space Berry curvature $B_{ad}(\bmr)$ does not affect the magnetic field $B(\bmr)$ on which our model Hamiltonian is based. To determine the quasiperiodicity of $\tilde{u}(\bmr)$ after the gauge fixing, we first consider the discrete translational symmetry $\Delta(\bmr + \bmR) = e^{i\tau^z\bmq_0\cdot\bmR} \Delta(\bmr) e^{-i\tau^z\bmq_0\cdot\bmR}$ for arbitrary lattice vector $\bmR$, which motivates writing
\begin{equation}
    \tilde{\bmu} (\bmr + \bmR) = e^{i\sigma^z\bmq_0\cdot\bmR + i\xi_\bmR(\bmr)} \tilde{\bmu}(\bmr),
    \label{eq:tildeu_periodicity}
\end{equation}
where $\xi_\bmR(\bmr)$ is some smooth phase to be determined. This leads to the quasiperiodic boundary condition in the Berry connection $A_{ad}(\bmr+\bmR) = A_{ad}(\bmr) + (\partial_x + i\partial_y)\xi_\bmR(\bmr)$. On the other hand, the far right-hand-side of Eq. (\ref{eq:A_ad}) gives $A_{ad}(\bmr+\bmR) = A_{ad}(\bmr) - (i/2)(R_x + iR_y)$, which indicates that $\xi_\bmR(\bmr) = \bmr\times\bmR/2 + \theta_\bmR$ where $\theta_\bmR$ is $\bmr$-independent. Self-consistency of Eq. (\ref{eq:tildeu_periodicity}) requires that for any two lattice vectors $\bmR_1$ and $\bmR_2$, $\xi_{\bmR_1 + \bmR_2}(\bmr) = \xi_{\bmR_2}(\bmr + \bmR_1) + \xi_{\bmR_1}(\bmr) + 2n\pi$ for some integer $n$, which means that $e^{i\theta_{\bmR_1+\bmR_2}} = e^{i(\theta_{\bmR_1} + \theta_{\bmR_2} + \bmR_1\times\bmR_2/2)}$. The general solution is $e^{i\theta_\bmR} = \eta_\bmR e^{i{\bm b}\cdot\bmR}$ for some constant vector $\bm b$. Similar analyses based on $C_2$ and $C_3$ symmetries fix $\bm b$ to reciprocal lattice vectors, which can then be set to 0. Ultimately we have
\begin{equation}
    \Tilde{\bmu}(\bmr + \bmR) = \eta_\bmR e^{i \left( \frac{\bmr\times\bmR}{2} + \sigma^z\bmq_0\cdot\bmR \right) } \Tilde{\bmu}(\bmr).
\end{equation}
Finally we define $\bmu(\bmr) = e^{-i\sigma^z\bmq_0\cdot\bmR} \Tilde{\bmu}(\bmr)$. Now combining $\bmu(\bmr)$ with Eq. (\ref{eq:twistedBloch}), we see that the transformed wave function $\phi_{n, \bmk s}(\bmr)$ defined by the first equality of Eq. (\ref{eq:eigenstate_mapped}) has the exact Bloch periodicity with momentum $\bmk$. In practice, the truncation of reciprocal lattice summation in Eq. (\ref{eq:eigenstate_mapped}) can slightly violate the orthonormality between bands. We solve this problem by performing Gram-Schmidt orthonormalization on the transformed LL wave functions $\phi_{\bmk s}^{nLL}(\bmr) = \psi_\bmk^{nLL}(\bmr) u_s(\bmr)$. To justify our algorithm, we have confirmed that the Berry curvature distribution $\Omega_\bmk$ and the quantum metric trace ${\rm tr}g_\bmk$ computed under the transformed plane-wave basis agree perfectly both among different choices of $(w, V, \psi)$ and with those computed from the original LL basis using the method described in Ref. \cite{AharonovCasher_TMD}.\\\\
We project un-screened Coulomb interactions to the highest band of the single-particle Hamiltonian Eq. \eqref{Adiabatic_Hamiltonian} in the main text, hence we will drop the band index from now on, and perform exact diagonalization on the many-body Hamiltonian
\begin{align}
    H=\sum_{{\bm k}} \epsilon_{{\bm k}}~ c^{\dagger}_{{\bm k}} c_{{\bm k}}+\frac{1}{2}\sum_{\substack{{\bm k}_i,{\bm k}_j\\{\bm k}_k,{\bm k}_l}} V_{ijkl} c^{\dagger}_{{\bm k}_i}c^{\dagger}_{{\bm k}_j}c_{{\bm k}_l} c_{{\bm k}_k},
\end{align}
where $c^{\dagger}_{{\bm k}} (c_{{\bm k}})$ creates (destroys) a spin-less hole with momentum ${\bm k}$; $ i,j,k,l$ are momentum labels, and $V_{ijkl}$ is a two-particle matrix element that we compute using the plane-wave basis constructed in Eq. \eqref{eq:eigenstate_mapped} and can be compactly written as
\begin{align}
    V_{ijkl}=\frac{1}{A}\left(\sum_{{\bm G}}\Lambda_{{\bm k}_1}^{{\bm q}+{\bm G}}\Lambda_{{\bm k}_2}^{-{\bm q}-{\bm G}} \frac{2\pi e^2}{\varepsilon (q+G)}\right).
    \label{Interaction_FormFactors}
\end{align}
Here, $q+G=|{\bm q}+{\bm G}|$ is the momentum transfer, $A$ is the area of the system and we make the standard choosing $V({\bm q}=0)=0$. The form factors in Eq. \eqref{Interaction_FormFactors} are defined as
\begin{align}
    \Lambda_{{\bm k}}^{{\bm q}+{\bm G}}=\sum_{{\bm G}',s}z^*_{{\bm k},{\bm G}',s}\,z_{{\bm k}+{\bm q},{\bm G}'+{\bm G},s},
    \label{FormFactorDef}
\end{align}
where the coefficients $z_{{\bm k},{\bm G},s}$ are obtained from Eq. \eqref{eq:eigenstate_mapped}. For all calculations presented in this work we have fixed the strength of the interaction scale to $U_{\text{int}}=e^2/\varepsilon \ell\approx 3.408~ \hbar\omega_c$. Furthermore, to ensure that our many-body results are are independent of the parameterization of $u_s({\bm r})$, we have performed the basis transformation given by Eq. \eqref{eq:eigenstate_mapped} with both $(w, V, \psi) = (18{\rm meV}, 9{\rm meV}, 128^\circ)$ \cite{devakul2021magic} and $(w, V, \psi) = (23.8{\rm meV}, 20.8{\rm meV}, 107.7^\circ)$ \cite{FCI_DiXiao} and confirmed that the resulting interaction matrix elements are identical. The only key requirement for the parameterization of $\bmu(\bmr)$ is that it winds over the unit cell.
%$\varepsilon=10$, which correspond to an interaction strength scale $e^2/\varepsilon \ell\approx 3.408~ \hbar\omega_c$, for a sample with twist angle $\theta = 1.67^\circ$ and lattice constant $a_0=0.3317$ nm \cite{devakul2021magic}. 
%

\section{Additional exact diagonalization results}

\begin{figure}
    \centering   \includegraphics[width=0.85\textwidth]{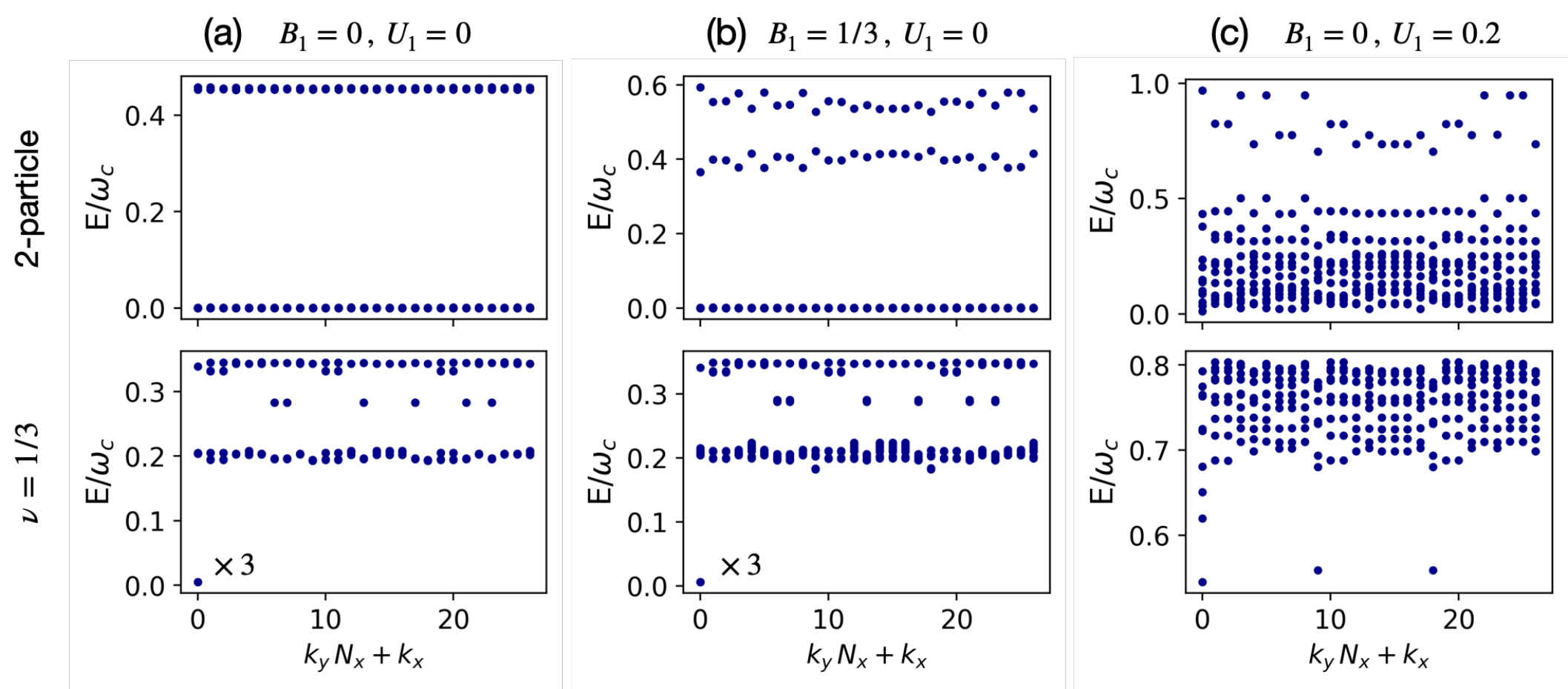}
    \caption{LLL {\it vs.} AC band {\it vs.} non-ideal band for first Haldane pseudopotential interactions. Two-particle spectrum for (a) the LLL; (b) an AC (ideal) band; and (c) a generic adiabatic band. Both (a) and (b) display a highly degenerate zero-energy manifold. Many-body spectrum at $\nu=1/3$ filling for (d) the LLL; (e) an AC (ideal) band; and (f) a generic adiabatic band. (d) and (e) display a three-fold degenerate ground zero-energy state in the same universality class as the Laughlin state, while (f) is in a CDW phase with non-zero ground state energy.}
    \label{fig:Haldane_ED}
\end{figure}

\begin{figure}
    \centering   \includegraphics[width=\textwidth]{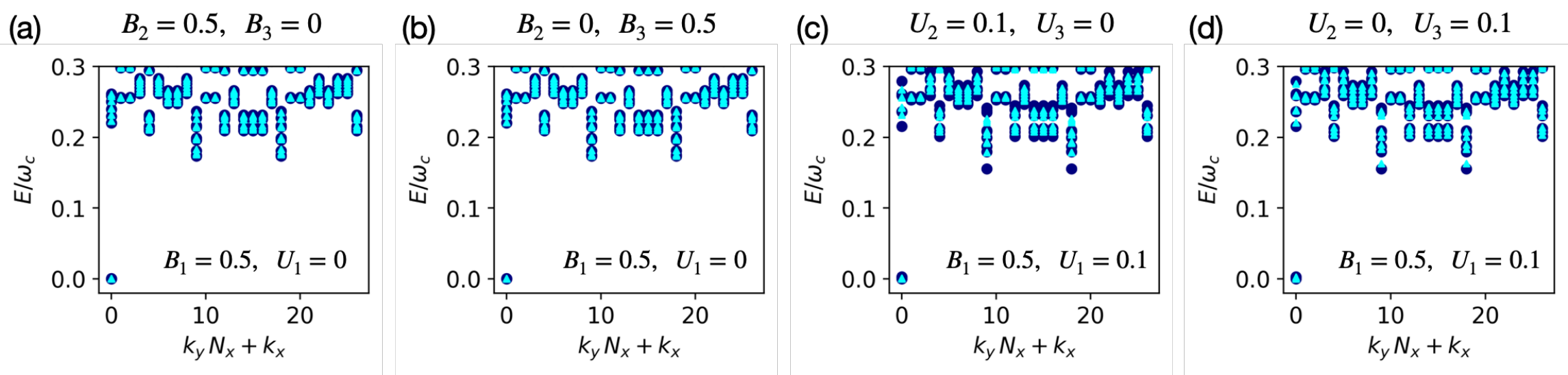}
    \caption{(a)-(b) Comparison between many-body spectra for AC bands with only first harmonic $B_1=0.5$ (dark blue) and for AC bands including higher harmonics (a) $B_2=0.5$ and (b) $B_3=0.5$ (cyan.) (c)-(d) Comparison between many-body spectra for adiabatic bands with $B_1=0.5$ and $U_1=0.1$ and adiabatic bands with higher harmonics (c) $U_2=0.1$ and (b) $U_3=0.1$ (cyan.)}
    \label{fig:Higher_Har}
\end{figure}

\begin{figure}
    \centering   \includegraphics[width=0.95\textwidth]{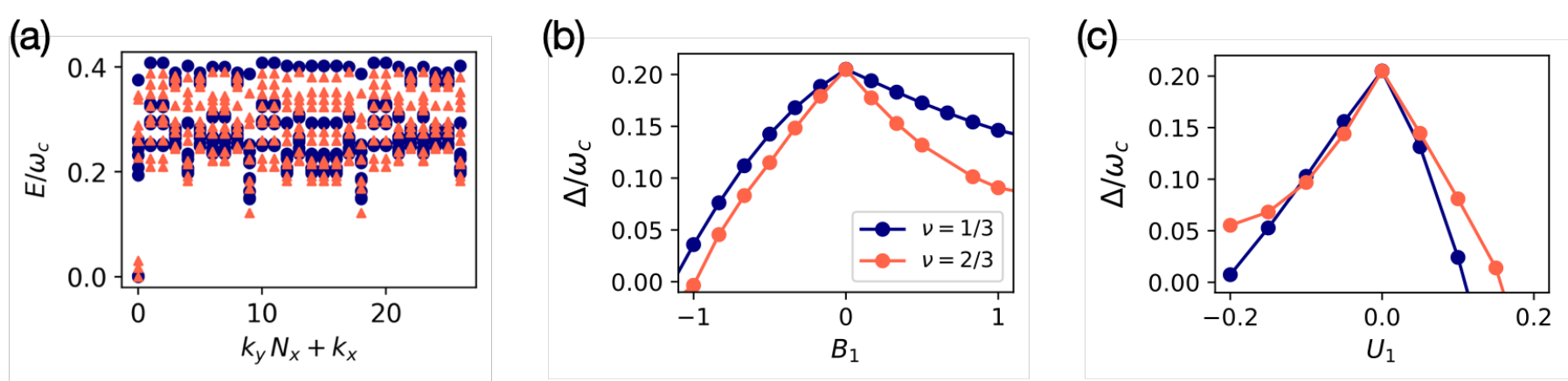}
    \caption{(a) Comparison of many-body spectra for $\nu=1/3$ and $\nu=2/3$ filling factors, for $B_1=1$. Both ground states are FCIs but the many-body gaps differ. (b) Evolution of the many-body gap $\Delta$ as a function of $B_1$, illustrating the particle-hole asymmetry due to the non-homogeneous Berry curvature. (c) Many-body gap for a band with $B_1=0$ as a function of $U_1$, which also shows particle-hole asymmetry. Labels are the same as in (b).}
    \label{fig:PH_Asymmetry}
\end{figure}

\begin{figure}
    \centering   \includegraphics[width=0.7\textwidth]{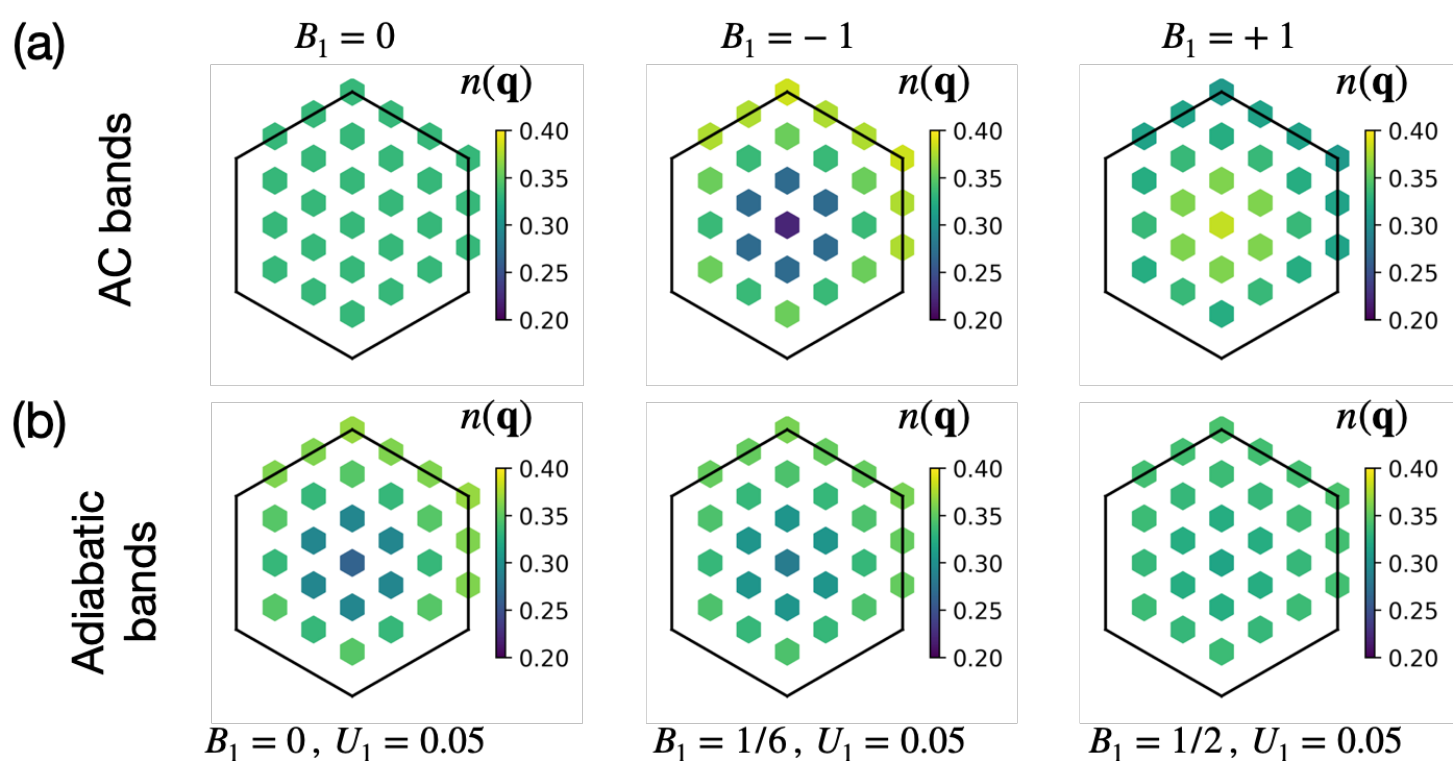}
    \caption{Momentum occupation plots for (a) different ideal bands and (b) different adiabatic bands. For the LLL the occupation is constant $n({\bm q})\approx 0.33$, but once $B_1\neq 0$ the occupation displays structure which is related to the Berry curvature distribution. For the non-ideal bands in the second row we see that the closest to constant occupation does not correspond to the band with $B_1=0$. The band with more homogeneous occupation is the one for which the many-body gap is larger. This illustrates the cancellation between contributions to the Berry curvature coming from $B({\bm r})$ and $U({\bm r})$ that we discuss in the main text.}
    \label{fig:Occupations}
\end{figure}

\begin{figure}
    \centering   \includegraphics[width=0.6\textwidth]{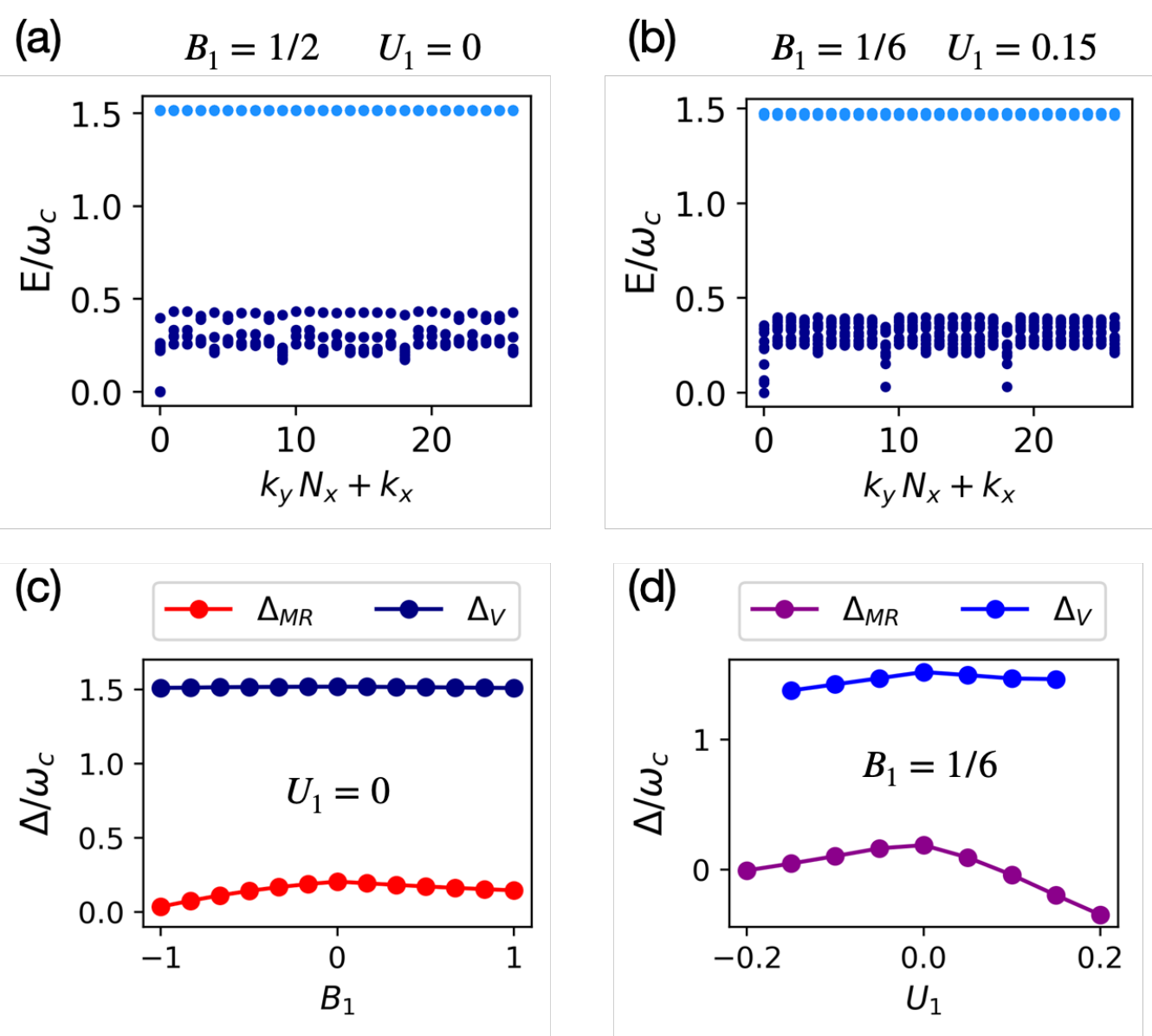}
    \caption{Exact diagonalization many-body spectra for (a) an AC band and (b) an adiabatic band. Dark blue corresponds to the fully-valley-polarized sector and light blue to the sector with one particle in the other valley. 10 eigenstates are shown for each spin-valley sector and model parameters are indicated. (c) Intra-valley gap $\Delta_{\text{MR}}$ and valley flip gap $\Delta_{\text{V}}$ for AC bands as a function of $B_1$. (d) Same as in (c) but for an adiabatic band as a function of $U_1$.}
    \label{fig:ValleyWaves}
\end{figure}

To further confirm the validity of our numerical approach, in Fig. \ref{fig:Haldane_ED}(a)-(c) we present many-body calculations for an idealized limit where the interaction consists only of the first Haldane pseudopotential, $V_1$, in a system with 27 unit cells. We compare the two-particle and the $\nu=1/3$ filling factor spectra for the (a) lowest Landau level ($B_1=U_1=0$), (b) an ideal AC band ($B_1\neq 0,~U_1=0$), and (c) an adiabatic band ($B_1= 0,~U_1\neq0$). As expected for the LLL, the two-particle spectrum displays a highly-degenerate zero-energy ground state manifold, separated from a two-fold degenerate finite-energy branch. For the AC band, which is ideal, the zero-energy manifold persists but the finite-energy branch now splits into two and acquires a dispersion. This behavior is expected due to the non-trivial quantum geometry of the AC band \cite{BergholtzMoessner}. In contrast, the two-particle spectrum for the non-ideal band is dense, as shown in Fig. \ref{fig:Haldane_ED}(c). This indicates that adding the residual potential to the AC Hamiltonian indeed destroys the ideal band structure.\\\\
From Fig. \ref{fig:Haldane_ED}(a)-(b) we see that the $\nu=1/3$ spectra for the LLL and the ideal band look quite similar: There is a three-fold degenerate ground state at many-body momentum ${\bm \gamma}$ well separated from all excitations, which is characteristic of a Laughlin-like state for this system size. This is in agreement with the LLL and the AC band having ideal quantum geometry. Finally, from the $\nu=1/3$ spectrum in Fig. \ref{fig:Haldane_ED}(c) we see that the ground state has finite energy and that the three-fold degeneracy at ${\bm \gamma}$ is lifted, with two states with many-body momentum ${\bm \kappa}$ and ${\bm \kappa^{\prime}}$ dropping down from the excited states to become nearly-degenerate with the lowest state at $\bm \gamma$. For the studied system size, this level crossing is typical of a transition from the FCI to a charge density wave ground state.\\\\ 
In this work we are mostly interested in situations where the $\nu=1/3$ many-body spectrum looks like in Figs. \ref{fig:Haldane_ED}(a)-(b). In such cases the ground state is topologically-ordered and three-fold quasi-degenerate. The small energy-splitting between the three ground states is caused by the finite system size, but it should become zero in the thermodynamic limit. We define
the many-body gap as the energy difference between the lowest-energy state that does not belong to the three-fold ground state manifold and the state with the highest energy within the ground state manifold. This is equivalent to taking the difference between the fourth and third lowest many-body energies if the system is in the FCI phase:
\begin{align}
    \Delta=\varepsilon_4-\varepsilon_3.
    \label{eq:MBGaps}
\end{align}
Here we will focus mainly on system sizes 27 and 36, where the three-fold topological ground state appears at the $\bm \gamma$ momentum sector (in agreement with \cite{Bernevig_Regnault_Rules}) and the lowest excitation energies emerge at the ${\bm \kappa}/{\bm \kappa^{\prime}}$ sectors. Eq. \eqref{eq:MBGaps} is what we plot in the main text figures to assess the stability of the FCI phase.\\\\
We now present some additional ED results, complimentary to the ones presented in the main text, using the un-screened Coulomb interaction, which complement the ones presented in the main text. In Fig. \ref{fig:Higher_Har} we show many-body spectra for different AC and adiabatic bands where we include only the first harmonics for the functions $B({\bm r})$ and $U({\bm r})$ and compare them with the spectra resulting from including higher-shell harmonics. The fact that the spectra in the two cases look almost identical justifies the approach we took of truncating the periodic functions that determine the adiabatic Hamiltonian to the first shell. In Fig. \ref{fig:PH_Asymmetry} we also compare many-body results for filling $\nu=1/3$ (considered in the main text) with filling $\nu=2/3$. In Fig. \ref{fig:Occupations} we show results for the many-body ground state momentum occupations for AC and adiabatic bands. We see that $n({\bm k})$ follows the distribution of the Berry curvature, with lower hole occupation in regions of peaked $\Omega_{\bm k}$. \\\\
In this work we focus on low-energy excitations within the fully-valley-polarized sector. However, when the valley degree of freedom is taken into account, a valley-flip excitation is another candidate for the lowest energy elementary excitation of the FCI. We have also performed ED calculations in the sector with one particle in the other valley and the results are shown in Fig. \ref{fig:ValleyWaves}. We see that the intra-valley gap is smaller than the inter-valley gap for the parameters considered in our work, which justifies restricting our analysis to the valley-polarized sector. This behavior has been observed in similar continuum models for moiré materials, e.g. Ref.~\cite{crepel2023anomalous}. \\\\

\section{Fourier expansions of some periodic functions }
In this work we have considered a magnetic field with only non-vanishing first-shell Fourier coefficients, $B_1$. The function $\chi({\bm r})$ that enters in the AC band wave function is related to the magnetic field by $\nabla^2\chi({\bm r})=B({\bm r})$, hence its Fourier coefficients are related to those of the magnetic field via $\chi_{\bm G}=-B_{\bm G}/|{\bm G}|^2$ and we can write
\begin{align}
    \chi({\bm r})=\sum_{\bm G}\chi_{\bm G}\,e^{i\,{\bm G}\cdot {\bm r}}=\sum_{\bm G}\left(-\frac{B_{1}}{G_0^2}\right)\,e^{i\,{\bm G}\cdot {\bm r}},
\end{align}
where we used that only the first-shell harmonics of the magnetic field are non-vanishing, and $G_0$ is the magnitude of the first-shell reciprocal lattice vectors. For all the derivations below we will take the harmonic $B_1$ as a perturbative parameter.\\\\
The periodic functions $e^{\chi(\bm r)}$ and $e^{2\chi(\bm r)}$ can be written as
\begin{align}
    e^{\chi({\bm r})}=1+\chi(\bm r)+\frac{1}{2}\chi^2(\bm r)+\cdots=\sum_{\bm G}\varphi_{\bm G}\, e^{i\,{\bm G}\cdot {\bm r}},\\
    e^{2\chi({\bm r})}=1+2\chi(\bm r)+2\chi^2(\bm r)+\cdots=\sum_{\bm G}\Phi_{\bm G}\, e^{i\,{\bm G}\cdot {\bm r}}.
\end{align}
The coefficients $\varphi_{\bm G}$ and $\Phi_{\bm G}$ can be obtained from the Fourier expansion of 
\begin{align}
    \chi^2(\bm r)=\sum_{{\bm G}, {\bm G}^{\prime}}\chi_{\bm G}\,\chi_{\bm G^{\prime}}\,e^{i({\bm G}+{\bm G}^{\prime})\cdot {\bm r}}=\sum_{\bm G} \left( \chi_{\bm G}\,\chi_{-\bm G}+2\chi_{\bm G}^2\, e^{i\,{\bm G}\cdot {\bm r}}\right)= 6\,\chi_1^2+2\sum_{\bm G\neq0}\chi_1^2\, e^{i\,{\bm G}\cdot {\bm r}},
\end{align}
which yields the following relations for the first two Fourier coefficients
\begin{align}
    \varphi_0&=1+3\,\chi_1^2,\qquad\qquad \varphi_1=\chi_1+\,\chi_1^2.\label{phiCoefficients}\\
    \Phi_0&=1+12\,\chi_1^2, \qquad \qquad\Phi_1=2\,\chi_1+4\,\chi_1^2.\label{PhiCoefficients}
\end{align}
Note that the functions $e^{\chi({\bm r})}$ and $e^{2\chi({\bm r})}$ have infinite non-vanishing Fourier coefficients, however given that $B_1\ll1$ we will truncate them to the first shell of harmonics. The AC many-body ground state wave functions involve products of $e^{\chi({\bm r}_i)}$ and $e^{2\chi({\bm r}_i)}$ over all $N$ particles composing the system. These products can be truncated to first order in $\chi_1$ as
\begin{align}
    \prod_{i=1}^Ne^{\chi({\bm r}_i)}&=\left(\sum_{\bm G}\varphi_{\bm G} e^{i\,{\bm G}\cdot {\bm r}_1}\right)\cdots \left(\sum_{\bm G}\varphi_{\bm G} e^{i\,{\bm G}\cdot {\bm r}_N}\right)\approx \varphi_0^N+\varphi_0^{N-1}\left(\sum_{{\bm G \neq 0}}\varphi_{1} e^{i\,{\bm G}\cdot {\bm r}_1}+\cdots+\sum_{\bm G\neq 0}\varphi_{1} e^{i\,{\bm G}\cdot {\bm r}_N}\right)+O(\varphi_{1}^2)\nonumber \\
    &\approx 1+\sum_{i=1}^N\sum_{\bm G \neq 0}\chi_1\,e^{i\,{\bm G}\cdot {\bm r}_i}+O(\chi_1^2)=1+\sum_{\bm G \neq 0}\chi_1 \rho_{\bm G}+O(\chi_1^2),
    \label{eq:expansion_X}
\end{align}
and
\begin{align}
    \prod_{i=1}^Ne^{2\chi({\bm r}_i)}&\approx 1+\sum_{i=1}^N\sum_{\bm G \neq 0} 2\chi_1\,e^{i\,{\bm G}\cdot {\bm r}_i}+O(\chi_1^2)=1+\sum_{\bm G \neq 0}2\chi_1 \rho_{\bm G}+O(\chi_1^2).
    \label{eq:expansion_2X}
\end{align}
\section{Berry curvature of the Aharonov-Casher band and energy of the adiabatic band}
\begin{figure}
    \centering   \includegraphics[width=0.9\textwidth]{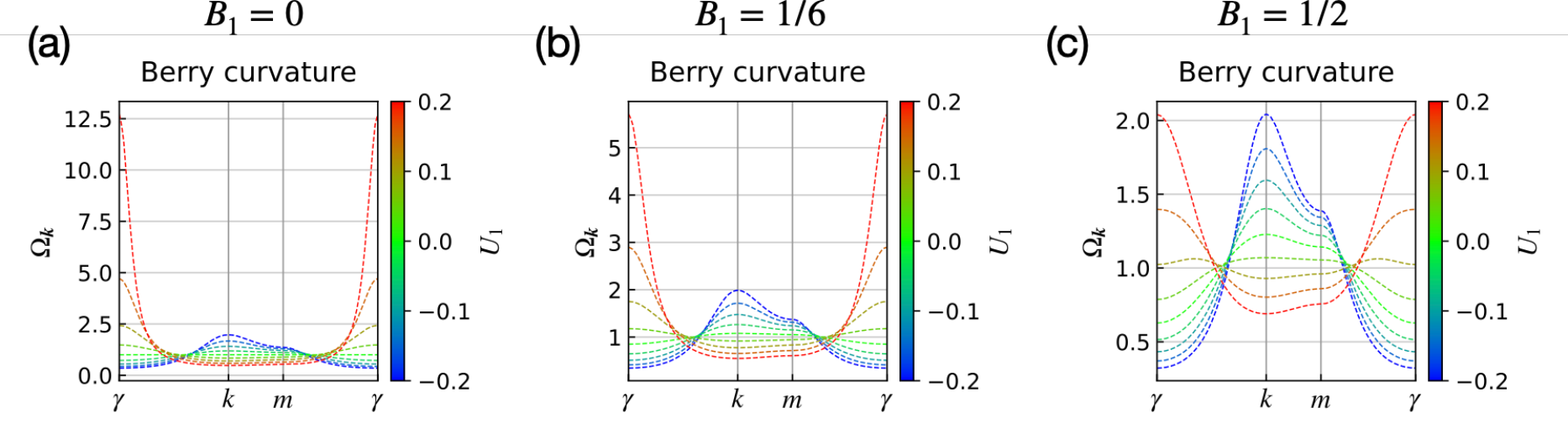}
    \caption{Berry curvatures for different adiabatic bands. (a) $\Omega_{\bm k}$ for the LLL plus a residual potential. We see that $U_1>0$ leads to a Berry curvature peak at $\bm \gamma$ and $U_1<0$ leads to Berry curvature peaks at ${\bm \kappa}/{\bm \kappa^{\prime}}$, which confirms our intuition obtained from Eq. \eqref{eq:Energy_perturb}. (b) and (c) show that a finite $B_1$ generically decreases the magnitude of the peaks of $\Omega_{\bm k}$.}
    \label{fig:Berry_Curvatures}
\end{figure}

From Eq. \eqref{Eq:Berry_Curvature} in the main text, we see that the Berry curvature of the AC band can be expressed in terms of the Fourier coefficients $\Phi_{\bm G}$
\begin{align}
    \Omega_{\bm k}&=\Omega_0+\frac{1}{2}\nabla^2_{\bm k}\ln\left( \sum_{\bm G} \Phi_{\bm G} \braket{\psi^{\text{LLL}}_{\bm k}|e^{i \,{\bm G}\cdot\hat{{\bm r}}}|\psi^{\text{LLL}}_{\bm k}}\right)=\Omega_0+\frac{1}{2}\nabla^2_{\bm k}\ln\left( \sum_{\bm G} \Phi_{\bm G}\overline{\eta}_{\bm G}\lambda_{\bm G}\,e^{i\ell^2{\bm k}\times {\bm G}}\right),
\end{align}
where the magnetic form factor is $\lambda_{\bm G}=e^{-|{\bm G}|^2\ell^2/4}$, $\overline{\eta}_{\bm G}=1$ if ${\bm G}/2$ belongs to the reciprocal lattice and $\overline{\eta}_{\bm G}=-1$ otherwise. If we keep only terms up to order $\chi_1$, we obtain 
\begin{align}
    \Omega_{\bm k}&\approx \Omega_0+\frac{1}{2}\nabla^2_{\bm k}\ln \left(1-\sum_{j=1}^62\chi_1\lambda_1\,e^{i{\bm k}\cdot{\bm R}_j}+O(\chi_1^2) \right)\approx \Omega_0+\frac{1}{2}\nabla_{\bm k}^2\left(-\sum_{j=1}^62\chi_1\lambda_1\,e^{i{\bm k}\cdot{\bm R}_j}+O(\chi_1^2)  \right)\nonumber \\
    &= \Omega_0-\frac{a^2B_1\lambda_1}{G_0^2}\sum_{j=1}^6\,e^{i{\bm k}\cdot{\bm R}_j} +O(B_1^2)=\Omega_0-\frac{4\pi\,\ell^2 B_1e^{-G_0^2\ell^2/4}}{\sqrt{3}G_0^2}\sum_{j=1}^6\,e^{i{\bm k}\cdot{\bm R}_j} +O(B_1^2).
\end{align}
In the previous expressions the vectors ${\bm R}_{j}$ belong to the first shell of lattice vectors. This is Eq. \eqref{eq:Berry_perturb} in the main text. From this linear expansion in $B_1$ we see that the Berry curvature is peaked at ${\bm \kappa}/{\bm \kappa^{\prime}}$ for $B_1>0$ and peaked at ${\bm \gamma}$ for $B_1<0$, as seen in Fig. 1(d) in the main text.\\\\
Once the residual potential is introduced, it induces a finite band width on the AC band. Focusing on the case $B_1=0$, we can estimate the corrections to the band energy introduced by $U(\bm r)$ as
\begin{align}
     \delta E_{\bm k}=\frac{\braket{\psi^{LLL}_{\bm k}|U(\bm r)|\psi^{LLL}_{\bm k}}}{\braket{\psi^{LLL}_{\bm k}|\psi^{LLL}_{\bm k}}}&=\braket{\psi_{\bm k}^{LLL}|\sum_{\bm G\neq 0}U_1e^{i{\bm G}\cdot {\bm r}}|\psi_{\bm k}^{LLL}}=-U_1e^{-G_0^2\ell^2/4}\sum_{j=1}^6 e^{i{\bm k}\cdot{\bm R}_j}.
     \label{eq:Energy_perturb}
\end{align}
We see that for $U_1>0$ the adiabatic band will have a single dip at ${\bm \gamma}$ and for $U_1<0$ there are two dips at ${\bm \kappa}/{\bm \kappa^{\prime}}$. Because we are working in the hole language, these dips will coincide with the peaks of the Berry curvature induced by $U({\bm r})$, because those are the regions where band mixing is more important. Fig. \ref{fig:Berry_Curvatures} confirms this behavior: $U_1>0$ corresponds to $\Omega_{\bm k}$ peaked at ${\bm \gamma}$, while $U_1<0$ will lead to $\Omega_{\bm k}$ peaked at ${\bm \kappa}/{\bm \kappa^{\prime}}$. From Eq. \eqref{eq:Energy_perturb} we expect that for a finite positive $B_1$ the Berry curvature is less pronounced than in the case of no periodic magnetic field, because the location of the maxima and minima induced by $B({\bm r})$ are opposite to those induced by $U({\bm r})$. This is also confirmed numerically in Fig. \ref{fig:Berry_Curvatures}. A similar argument applies for the combination of parameters $B_1<0$ and $U_1<0$.

\section{Approximation to the magnetoroton gap}
\begin{figure}
    \centering   \includegraphics[width=0.9\textwidth]{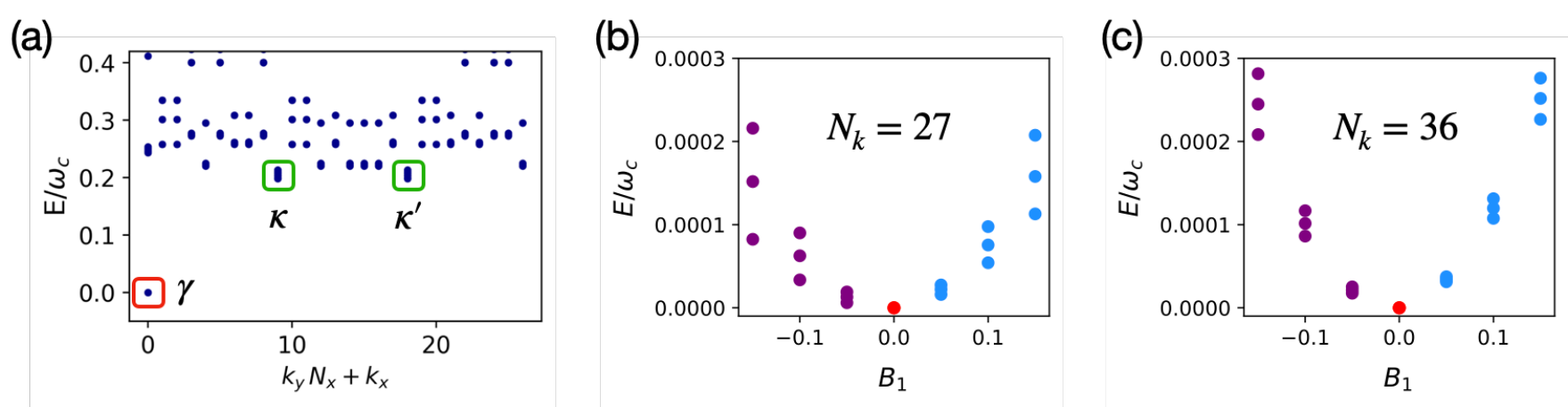}
    \caption{(a) Many-body spectrum for an AC band with $B_1=0.1$ and $N_k=27$ unit cells. The three quasi-degenerate topological ground states are at $\bm \gamma$ and are indicated in red. The lowest excitations are located at ${\bm \kappa}/{\bm \kappa^{\prime}}$ and are indicated in green. (b)-(c) Evolution of the three topological ground states as a function of $B_1$ for systems with 27 and 36 unit cells, respectively. Note that for $B_1\neq0$ the energy change with respect to the LLL ground state energy is $10^{3}$ times smaller than the many-body gap. Note that for a given $B_1$, the ground state splitting decreases with system size, indicating that the three states are degenerate in the thermodynamic limit.}
    \label{fig:GS_Splitting}
\end{figure}

\begin{figure}
    \centering   \includegraphics[width=0.95\textwidth]{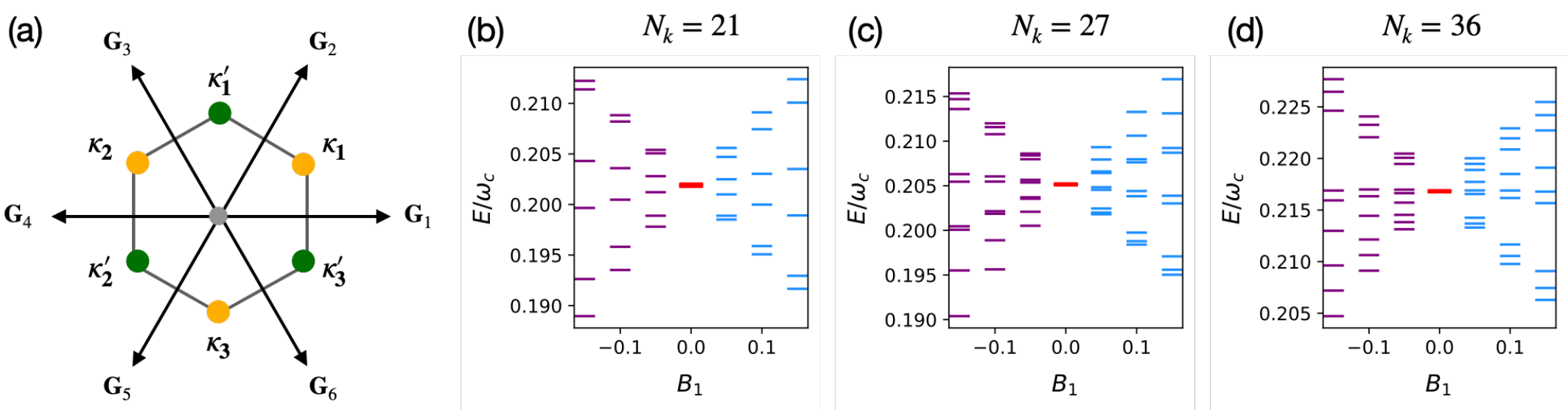}
    \caption{(a) Schematic illustration of our minimal model for the magnetorotons at the corners of the Brillouin zone. The $\bm \kappa$ and $\bm \kappa'$ points are decoupled and degenerate in energy. (b),(c),(d) Energy of excitations at the $\bm \kappa$-point, with respect to the ground state energy, for different system sizes and as a function of $B_1$. For the system with 21 unit cells the three topological ground states are in different momentum sectors, hence there are only 6 low-energy magnetorotons at $\bm \kappa$. For system sizes 27 and 36, the three ground states are in the same sector, resulting into 9 magnetoroton excitations at $\bm \kappa$. The lowest excitation has consistently lower energy for $B_1<0$ than for $B_1>0$ across all system sizes.}
    \label{fig:kappa_excitations}
\end{figure}

\begin{figure}
    \centering   \includegraphics[width=0.7\textwidth]{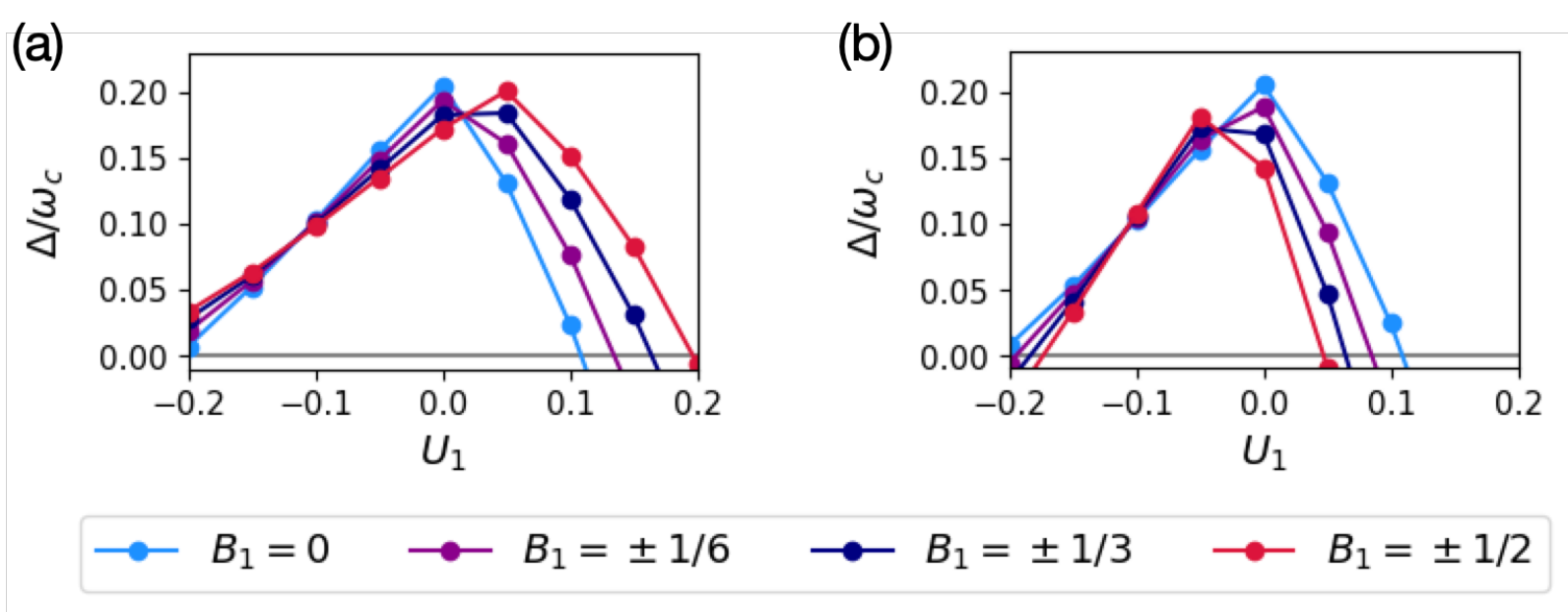}
    \caption{(a) Magnetoroton gap as a function of $U_1$ for different values of $B_1>0$. The maximum gap is attained for a finite $U_1>0$. (b) Magnetoroton gap as a function of $U_1$ for different values of $B_1<0$. The maximum gap is attained for a finite $U_1<0$.}
    \label{fig:MRGap_Linecut}
\end{figure}

In this section we will focus on adiabatic
bands that are perturbatively close to the LLL, namely bands where the oscillating pieces of the magnetic field and the residual potential are small in magnitude, {\it i.e.}, $|B_1|, |U_1|<<1$ (remember that $B_1$ has units of $1/\ell^2$ and $U_1$ has units of $\omega_c$.) We are interested in calculating the $\nu=1/3-$filling magnetoroton gap
\begin{align}
    \Delta^{ad}({\bm q})=E^{ad}({\bm q})-E_0^{ad},
    \label{eq:AC_MRGap}
\end{align}
where $E_0^{ad}$ is the many-body ground state energy and $E^{ad}({\bm q})$ is the energy of the lowest neutral excitation at momentum $\bm q$. In particular we want to compare Eq. \eqref{eq:AC_MRGap} with the LLL magnetoroton gap
\begin{align}
    \Delta^{LLL}({\bm q})=E^{LLL}({\bm q})-E_0^{LLL},
\end{align}
where $E_0^{LLL}$ is the many-body ground state energy of the LLL and $E^{LLL}({\bm q})$ the magnetoroton minimum energy. \\\\
The ED calculations presented here are performed by applying periodic boundary conditions, hence the states live on a torus. This results in a three-fold quasi-degenerate topological ground state, as can be seen in Fig. \ref{fig:GS_Splitting}. We know that in the thermodynamic limit, due to continuous translational symmetry, a non-trivial $B(\bm r)$ and $U(\bm r)$ will not mix different ground states, and that the energy shift due to the coupling between ground state and excited states that differ by a reciprocal lattice vector will be quadratic in $B_1$ and $U_1$, to lowest order. This is partially numerically verified by Figs. \ref{fig:GS_Splitting} (b), (c), where we show the splitting between the three ground states as a function of $B_1$. Because, as we will show below, the main correction that we identify for the excited states is linear in $B_1$ and $U_1$, we will not consider the quadratic corrections to any state from now on.\\\\
We now analyze the perturbation effects of $B(\bm r)$ and $U(\bm r)$ on the magnetoroton modes of the LLL, which we describe in the {\it single mode approximation} (SMA) \cite{Girvin_MacDonald_Platzman_1,Girvin_MacDonald_Platzman_2},
\begin{equation}
    \Ket{\Psi_{\bm q}^{LLL}} = \frac{1}{\sqrt{N}} \overline \rho_{\bm q} \Ket{\Psi_0^{LLL}},
\end{equation}
where $N$ is the total particle number, $\overline\rho_{\bm q}$ is the LLL-projected density operator and $\Ket{\Psi_0^{LLL}}$ is a ground state. The condition of one flux quantum per unit cell, $2\pi\ell^2=A_{\text{UC}}$, results in the minimum of the LLL magnetoroton excitation being approximately at the edges of the Brillouin zone defined by the periodicity of the magnetic field. We note that even for small $B_1$, the lowest excitations are at ${\bm \kappa}$ and ${\bm \kappa'}$, as can be seen in Fig. \ref{fig:GS_Splitting} (a). These excitations are related via $C_2-$rotation but are not coupled by a reciprocal lattice vector. Hence, we limit our analysis to the space spanned by the three equivalent Brillouin zone corners, ${\bm q} = {\bm\kappa}_1$, ${\bm\kappa}_2$ and ${\bm\kappa}_3$, which are illustrated in Fig. \ref{fig:kappa_excitations} (a).
The single mode approximation at $\bm \kappa_i$ can be built on top of each topological ground state, which means that the full set of low-energy neutral excitations at $\bm \kappa_i$ consists of nine states, three excitations for each ground state. If there were no coupling between the neutral excitations obtained from each topological ground state, we would observe three sets of three-fold degenerate excitations and this is what we expect to happen in the thermodynamic limit. However, the coupling between different neutral excitations is non-vanishing for finite size systems, resulting in splittings between the nine energy levels at $\bm \kappa/\bm \kappa'$. In Fig. \ref{fig:kappa_excitations} we plot the energies of the lowest nine (or six) many-body states at $\bm \kappa$, the corner of the Brillouin zone, as a function of $B_1$ and for different system sizes. We see that the splitting is almost negligible at $B_1=0$.\\\\
The analytical tractability of AC bands allows us to write down the magnetoroton states at momentum ${\bm q}$ in the SMA as
\begin{align}
    \ket{\Psi_{\bm q}^{AC}}=e^{X(\{\bm r_i\})}\ket{\Psi_{\bm q}^{LLL}}=\frac{1}{\sqrt{N}}e^X\overline{\rho}_{\bm q}\ket{\Psi_0^{LLL}},
    \label{SMA1}
\end{align}
where $X = \sum_{i=1}^N\chi({\bm r}_i)$ with function $\chi(\bm r)$ satisfying $\nabla^2\chi = B(\bm r)$. There is an alternative definition,
\begin{align}
    \ket{\widetilde{\Psi}_{\bm q}^{AC}}=\frac{1}{\sqrt{N}}\widetilde{\rho}_{\bm q}\ket{\Psi_0^{AC}}=\frac{1}{\sqrt{N}}\widetilde{\rho}_{\bm q}\,e^X\ket{\Psi_0^{LLL}}\JS{,}
    \label{SMA2}
\end{align}
where $\widetilde{\rho}_{\bm q}$ is the AC band-projected density operator.
Eqs. \eqref{SMA1} and \eqref{SMA2} span the same Hilbert space and for simplicity we will work with the basis defined in Eq. \eqref{SMA1}.\\\\
One can note that states $\ket{\Psi_{\bm q}^{AC}}$ and $\ket{\Psi_{\bm k}^{AC}}$ are not necessarily orthogonal, as they can be coupled by a reciprocal lattice vector. This implies that the structure factor is not purely diagonal in a basis of reciprocal lattice vectors, namely
\begin{align}
    \braket{\Psi_{\bm k}^{AC}|\Psi_{\bm q}^{AC}}&=\frac{1}{N}\braket{\overline{\rho}_{-{\bm k}}\,e^{2X}\,\overline{\rho}_{{\bm q}}}_0^{LLL}\approx\frac{1}{N}\left(  \braket{\overline{\rho}_{-{\bm k}}\,\overline{\rho}_{{\bm q}}}_0^{LLL}\delta_{{\bm k},{\bm q}}+2\sum_{\bm G \neq 0}\chi_{1}\braket{\overline{\rho}_{-{\bm k}}\overline{\rho}_{{\bm G}}\overline{\rho}_{{\bm q}}}_0^{LLL}\delta_{{\bm k},{\bm q}+{\bm G}}+O(\chi_{1}^2)\right),
    \label{StructureFactorMat}
\end{align}
where we have used the expansion in Eq. \eqref{eq:expansion_2X}. The lowest-order off-diagonal elements of Eq. \eqref{StructureFactorMat} are the three-point density correlation functions of the LLL ground state. Similarly, the matrix elements of the residual potential are given by
\begin{align}
    \braket{\Psi_{\bm k}^{AC}|U|\Psi_{\bm q}^{AC}} = \frac{1}{N}\braket{\overline\rho_{-\bm k} e^{X} U e^{X} \overline\rho_{\bm q}}_0^{LLL} \approx \frac{1}{N}\braket{\overline\rho_{-\bm k} U \overline\rho_{\bm q}}_0^{LLL} + O(\chi_1 U_1)
    = \frac{1}{N} \sum_{\bm G\ne 0} U_1 \braket{\overline\rho_{-\bm k} \overline\rho_{\bm G} \overline\rho_{\bm q}}_0^{LLL} + O(\chi_1U_1),
\end{align}
and the matrix elements of the interacting Hamiltonian are given by
\begin{align}
    \braket{\Psi_{\bm k}^{AC}|H_{\text{int}}|\Psi_{\bm q}^{AC}}&=\frac{1}{N}\braket{\overline{\rho}_{-{\bm k}}\, e^X \,H_{\text{int}}\,e^X\,\overline{\rho}_{{\bm q}}}_0^{LLL}\label{InteractionMat} \\
    \approx&\frac{1}{N}\left( \braket{\overline{\rho}_{-{\bm k}}\, H_{\text{int}}\,\overline{\rho}_{{\bm q}}}_0^{LLL}\delta_{{\bm k},{\bm q}}+\sum_{\bm G\neq 0}\chi_{1}(\braket{\overline{\rho}_{-{\bm k}}\,\overline{\rho}_{{\bm G}} H_{\text{int}}\,\overline{\rho}_{{\bm q}}}_0^{LLL}+\braket{\overline{\rho}_{-{\bm k}}\, H_{\text{int}}\overline{\rho}_{{\bm G}}\,\overline{\rho}_{{\bm q}}}_0^{LLL})\delta_{{\bm k},{\bm q}+{\bm G}}+O(\chi_{1}^2)\right) \nonumber,
\end{align}
where we have used Eq. \eqref{eq:expansion_X} for the truncation in the second line. We see that the off-diagonal elements of the interaction Hamiltonian generally consist of expectation values of five density operators in the LLL ground state. The energy of the neutral excitation for a given momentum $\bm q$ in the Brillouin zone, $E^{ad}({\bm q})$, can be estimated by solving the generalized eigenvalue problem
\begin{align}
    {\sum_{\bm G}}\braket{\Psi_{{\bm q}+{\bm G'}}^{AC}|H_{\text{int}}+U|\Psi_{{\bm q}+{\bm G}}^{AC}}\,u_{\bm q,\bm G}=E^{ad}({\bm q})\sum_{\bm G}\braket{\Psi_{{\bm q}+{\bm G'}}^{AC}|\Psi_{{\bm q}+{\bm G}}^{AC}}\,u_{\bm q ,\bm G}\,,
    \label{GeneralizedEVproblem1}
\end{align}
where $u_{\bm q, \bm G}$ is a generalized eigenvector. Even with the truncations obtained in Eqs. \eqref{StructureFactorMat}-\eqref{InteractionMat}, solving the generalized eigenvalue problem would require a numerical calculation of several $n$-point correlation functions of the Laughlin state. \\\\
As we argued before, when $B_1 \ne 0$, we can focus on only the three $\bm \kappa$ points in the Brillouin zone, that we denote by ${\bm \kappa}_i$, with $i=1,2,3$ and are related by $C_3-$symmetry. The eigenvalue problem truncated to these three points reduces to
\begin{align}
    \begin{pmatrix}
        \overline{V}_0({\bm \kappa_1}) &
        \overline{H}_{1\,{\bm G_0},{\bm G_4}}^{~{\bm \kappa_1}} &
        \overline{H}_{1\,{\bm G_0},{\bm G_5}}^{~{\bm \kappa_1}} \\
        \overline{H}_{1\,{\bm G_4},{\bm G_0}}^{~{\bm \kappa_1}} &
        \overline{V}_0({\bm \kappa_2}) &
        \overline{H}_{1\,{\bm G_4},{\bm G_5}}^{~{\bm \kappa_1}} \\
        \overline{H}_{1\,{\bm G_5},{\bm G_0}}^{~{\bm \kappa_1}} &
        \overline{H}_{1\,{\bm G_5},{\bm G_4}}^{~{\bm \kappa_1}} &
        \overline{V}_0({\bm \kappa_3})
    \end{pmatrix}
    \,u_{\bm \kappa_1}
    =E^{ad}({\bm \kappa_1})\begin{pmatrix}
        \overline{S}_0({\bm \kappa_1})&\chi_1\,\overline{S}_{1\,{\bm G_0},{\bm G_4}}^{~{\bm \kappa_1}}&\chi_1\,\overline{S}_{1\,{\bm G_0},{\bm G_5}}^{~{\bm \kappa_1}}\\
        \chi_1\,\overline{S}_{1\,{\bm G_4},{\bm G_0}}^{~{\bm \kappa_1}}&\overline{S}_0({\bm \kappa_2})&\chi_1\,\overline{S}_{1\,{\bm G_4},{\bm G_5}}^{~{\bm \kappa_1}}\\
        \chi_1\,\overline{S}_{1\,{\bm G_5},{\bm G_0}}^{~{\bm \kappa_1}}&\chi_1\,\overline{S}_{1\,{\bm G_5},{\bm G_4}}^{~{\bm \kappa_1}}&\overline{S}_0({\bm \kappa_3})
    \end{pmatrix}\,u_{\bm \kappa_1}.
    \label{GeneralizedEVproblem2}
\end{align}
The reciprocal lattice vectors ${\bm G}_i$ are indicated in Fig. \ref{fig:kappa_excitations}(a)% below
, $\overline{S}_0({\bm k})= \braket{\overline{\rho}_{-{\bm k}}\,\overline{\rho}_{{\bm k}}}_0^{LLL}/N$ is the projected structure factor and $\overline{V}_0({\bm k})=\braket{\overline{\rho}_{-{\bm k}}\, H_{\text{int}}\,\overline{\rho}_{{\bm k}}}_0^{LLL}/N$ is an interaction matrix element calculated on the LLL ground state, which due to the $C_3-$symmetry are all equal, and for shorthand we write them as $S_0\equiv\overline{S}_0({\bm \kappa_i})$ and $\overline{V}_0\equiv V_0({\bm \kappa_i})$, with $i=1,2,3$. We have also defined 
\begin{align}
    \overline{H}_{1\,{\bm G}, {\bm G}'}^{~{\bm k}} &= \chi_1 \overline{V}_{1\,{\bm G}, {\bm G}'}^{~{\bm k}} + \frac{U_1}{2} \overline{S}_{1\,{\bm G}, {\bm G}'}^{~{\bm k}}, \\
    \overline{V}_{1\,{\bm G}, {\bm G}'}^{~{\bm k}}&=\frac{1}{N}\sum_{\widetilde{\bm G}}(\braket{\overline{\rho}_{-{\bm k}-{\bm G}}\,\overline{\rho}_{\widetilde{\bm G}} H_{\text{int}}\,\overline{\rho}_{{\bm k}+{\bm G'}}}_0^{LLL}+\braket{\overline{\rho}_{-{\bm k}-{\bm G}}\, H_{\text{int}}\overline{\rho}_{\widetilde{\bm G}}\,\overline{\rho}_{{\bm k}+{\bm G'}}}_0^{LLL})\equiv V_1,\\
    \overline{S}_{1\, {\bm G},{\bm G'}}^{~{\bm k}}&=\frac{2}{N}\sum_{\widetilde{\bm G}}\braket{\overline{\rho}_{-{\bm k}-{\bm G}}\overline{\rho}_{\widetilde{\bm G}}\overline{\rho}_{{\bm k}+{\bm G'}}}_0^{LLL}\equiv S_1.
\end{align}
The off-diagonal elements in Eq. \eqref{GeneralizedEVproblem2} can be related by $C_3-$symmetry, which allows us to write the problem in a more compact form,
\begin{align}
    \begin{pmatrix}
        V_0 & \chi_1V_1^* + U_1S_1^*/2 & \chi_1V_1 + U_1S_1/2 \\
        \chi_1V_1 + U_1S_1/2 & V_0 & \chi_1V_1^* + U_1S_1^*/2 \\
        \chi_1V_1^* + U_1S_1^*/2 & \chi_1V_1 + U_1S_1/2 & V_0
    \end{pmatrix} u_{\bm \kappa_1}=E^{ad}({\bm \kappa_1})\begin{pmatrix}
        S_0 & \chi_1S_1^* & \chi_1S_1 \\
        \chi_1S_1 & S_0 & \chi_1S_1^* \\
        \chi_1S_1^* & \chi_1S_1 & S_0
    \end{pmatrix}u_{\bm \kappa_1},
    \label{eq:3x3matrixproblem}
\end{align}
where $V_0$, $S_0$ are real constants and $V_1$, $S_1$ are complex constants. The eigenvalues of this problem are
\begin{equation}
    E_j^{ad}(\bm \kappa_1) = \frac{V_0 + {\rm Re} \left[ e^{i\frac{2\pi j}{3}} (2\chi_1 V_1 + U_1S_1) \right]}{S_0 + 2\chi_1 {\rm Re} \left( e^{i\frac{2\pi j}{3}} S_1 \right)},
\end{equation}
where $j = 0, \pm 1$. Taking the perturbative limit of both $|\chi_1|\ll1$ and $|U_1|\ll 1$, we obtain 
\begin{equation}
    E_j^{ad}(\bm\kappa_1) = \frac{V_0}{S_0} + \frac{1}{S_0} {\rm Re} \left\{ e^{i\frac{2\pi j}{3}} \left[ -\frac{2B_1}{G_0^2} \left( V_1 - \frac{V_0S_1}{S_0} \right) + U_1S_1 \right] \right\} = \frac{V_0}{S_0} + {\rm Re} \left[ e^{i\frac{2\pi j}{3}} (\tilde a B_1 + \tilde c U_1) \right],
    \label{eq:Ej_compact}
\end{equation}
where we have used $B_1 = -G_0^2\,\chi_1$ and defined the complex constants
\begin{align}
    \tilde a = -\frac{2}{G_0^2S_0}\left(\frac{V_0S_1}{S_0} - V_1\right), \quad\quad\text{and}\quad\quad \tilde c = \frac{S_1}{S_0}. 
\end{align}
The specific values of $\tilde a$ and $\tilde c$ can be computed, for instance, via Monte Carlo simulations in the plasma analogy, but we can infer some constraints on them from our numerical results. From Fig. \ref{fig:Trace_Deviations} of the main text we see a line across the $B_1$--$U_1$ space along which the many-body gap is maximal and almost keeps constant. In our linearized theory, this is only possible if $\tilde a/\tilde c$ is a real number (or equivalently, $V_1/S_1$ is a real number), and in that case the line of maximal gap is given by $U_1/B_1 = -\tilde a/\tilde c$. The positive slope of the line indicates that $\tilde a/\tilde c < 0$.\\\\
By taking $U_1 = 0$ in Eq. (\ref{eq:Ej_compact}) the AC limit is recovered. Our model qualitatively captures the fact that the energy of the lowest neutral excitation at $\bm \kappa$ is smaller for an AC band with $B_1\neq0$ than for the LLL, as observed in Fig. \ref{fig:Berry_Fluctuations}(c) in the main text. Eq. (\ref{eq:Ej_compact}) also explains the asymmetric slopes in the many-body gap on positive and negative values of $B_1$ observed in Fig. \ref{fig:Berry_Fluctuations} (c) in the main text, where
\begin{equation}
    \alpha = \max_{j = 0, \pm 1} {\rm Re} \left( e^{i\frac{2\pi j}{3}} \tilde a \right), \quad\quad\text{and}\quad\quad \beta = -\min_{j = 0, \pm 1} {\rm Re} \left( e^{i\frac{2\pi j}{3}} \tilde a \right),
\end{equation}
are the corrections entering Eq. \eqref{eq:AC_magnetorotongap} in the main text. Moreover, our linearlized theory also explains the behavior of the many-body gap away from the ideal and flat band limit, by incorporating the effects of $U_1$. In Fig. \ref{fig:MRGap_Linecut}, we plot line-cuts for the many-body gap for fixed values of $B_1$ as a function of $U_1$, where it is clearly seen that the maximal gap happens at a finite $U_1>0$ for $B_1>0$ and at $U_1<0$ for $B_1<0$. Additionally, regardless of the sign of $B_1$, the slopes of the many-body gap are asymmetric on positive and negative sides of $U_1$, which can be explained by Eq. \eqref{eq:Ej_compact}.

\end{document}